\documentstyle [12pt,epsfig]{article} 
\makeatletter

\@addtoreset{equation}{section}
\makeatother

\newcommand{\ba}{\begin{eqnarray}}
\newcommand{\ea}{\end{eqnarray}}
\newcommand{\beq}{\begin{equation}}
\newcommand{\eeq}{\end{equation}}

\newcommand{\RC}{{\rm C}}

\newcommand{\rx}{{\rm x}}
\newcommand{\rt}{{\rm t}}
\newcommand{\ry}{{\rm y}}
\newcommand{\rz}{{\rm z}}
\newcommand{\rs}{{\rm s}}
 
\begin{document}
\newcommand{\BS}{\bigskip}
\newcommand{\SECTION}[1]{\BS{\large\section{\bf #1}}}
\newcommand{\SUBSECTION}[1]{\BS{\large\subsection{\bf #1}}}
\newcommand{\SUBSUBSECTION}[1]{\BS{\large\subsubsection{\bf #1}}}

\begin{titlepage}
\begin{center}
\vspace*{2cm}
{\large \bf Langevin's `Twin Paradox' paper revisited}  
\vspace*{1.5cm}
\end{center}
\begin{center}
{\bf J.H.Field }
\end{center}
\begin{center}
{ 
D\'{e}partement de Physique Nucl\'{e}aire et Corpusculaire
 Universit\'{e} de Gen\`{e}ve . 24, quai Ernest-Ansermet
 CH-1211 Gen\`{e}ve 4.
}
\newline
\newline
   E-mail: john.field@cern.ch
\end{center}
\vspace*{2cm}
\begin{abstract}
   An in-depth and mathematically-detailed analysis of Langevin's popular 1911
     article on the special theory of relativity is presented. For the reader's
    convenience, English translations of large parts of the original French text 
    are given. The self-contradictory nature of many of Langevin's assertions is pointed out.
   Of special interest is the analysis of the exchange
    of light signals between the travelling and stay-at-home twins in Langevin's
    thought experiment, in which antinomies are found in the conventional relativistic treatment.
     Their resolution shows that the physical basis
        of the differential aging effect in the experiment is
        not `length contraction', as in the conventional interpretation,
       but instead the application of the correct relative velocity transformation
       formula. The spurious nature of the
       correlated `length contraction' and `relativity of simultaneity' effects
        of conventional special relativity is also demonstrated.
       In consequence, an argument given, claiming to demonstrate that an upper limit of $c$ on
       the speed of any physical signal
      is required by causality, is invalid. Its conclusion is  
      also in contradiction with astronomical observations and 
      the results of a recent experiment.

 \par \underline{PACS 03.30.+p}
\vspace*{1cm}
\end{abstract}
\end{titlepage}
 
\SECTION{\bf{Introduction}}
    The purpose of the present paper is an in-depth re-examination of Langevin's 1911 popular
  exposition of special relativity~\cite{Langevin}, containing the famous travelling-twin thought experiment,
   in the light of a series of recent papers~\cite{JHFLLT,JHFSTP1,JHFSTP2,JHFSTP3} on space-time physics by the
   present author.
   It is concluded in these papers that important mathematical and conceptual errors exist in standard
    text book treatments
   of special relativity. The nature of these errors will become apparent in the discussion below of the
   special-relativistic effects presented in Ref.~\cite{Langevin}. The physical basis of the differential
    aging effect in the travelling-twin experiment was previously discussed in Ref.~\cite{JHFSTP3}. In the 
   present paper, particular consideration will also be given to the exchange of light signals, between
   the Earth-bound and travelling twins, considered by Langevin. This analysis reveals new antinomies in
    conventional special relativity theory that are resolved by the arguments
    given in Refs.~\cite{JHFLLT,JHFSTP1,JHFSTP2,JHFSTP3}, and recalled in the present paper.
   \par The early parts of Ref.~\cite{Langevin} compare and contrast classical mechanics and
    classical electromagnetism as they were understood at the beginning of the 20th Century (i.e. before
    the advent of quantum mechanics). The standard space-time `effects' of special relativity:
      `Relativity of Simultaneity' (RS), `Length Contraction' (LC) and Time Dilatation (TD)\footnote{Throughout the
     present paper the spurious RS and LC effects, when mentioned without the use of their 
    acronyms, are given in quotation marks to distingish them from the physical and experimentally
     verified TD effect.}
    are presented in a pedagogical account, in which the concepts of space-like and time-like separated
     event pairs are also introduced.
    In the second part of Ref.~\cite{Langevin} the travelling-twin thought experiment is introduced and discussed.
    \par The present paper aims to present a detailed mathematical analysis of
    the special relativistic effects ---in particular RS, LC and TD and differential aging in the
    travelling-twin experiment--- that are verbally presented in Langevin's paper, which contains
     no equations. The analysis of RS, LC and TD is presented in Section 2, that of the thought
     experiment in Section 3. Both of these sections contain English translations of the 
    relevant passages of the French text of Ref.~\cite{Langevin}, before the following critical
     discussions. The remainder of the present section contains,
    in the guise of a general introduction,  a pr\'{e}cis, with English translations of
   four key passages, of the early parts of Ref.~\cite{Langevin} which shed light on
     Langevin's intellectual attitude in his subsequent discussion of space and time in
     special relativity. The final Section 4 contains a summary of the
     conclusions of the previous sections and some closing remarks.
     \par Langevin contrasts Newtonian concepts of space and time, described by the Galilean transformations
       $x' = x-vt$, $y' = y$, $z'= z$ and $t' = t$, which are asserted to be required by the laws
       of classical mechanics, with the new concepts required by the laws of electromagnetism\footnote{`Classical
        electromagnetism' in modern parlance, since quantum mechanics was yet to be discovered in 1911.}.
         In particular, the Lorentz transformation, which leaves invariant the form of
        Maxwell's equations in different inertial frames, requires the Newtonian space-time concepts
        to be modified. Langevin, in concordance with the title of Ref.~\cite{Langevin}: L'\'{e}volution de l'espace
        et du temps', introduces the idea of `natural selection' of physical theories analogous
       to the conflict and symbiosis of different living species leading to their Darwinian evolution.
       Langevin however viewed the then-current situation of classical mechanics and electromagnetism
       more as one of unresolved conflict than of symbiosis:
 \begin{quotation}
\textwidth 15cm      
        Electromagnetism is as well adapted to its essential domain, as classical mechanics is to
            its. With its very special concept of a medium which transmits, by contact interactions, electric
            and magnetic fields which  describe the state of this medium, and the special form it imposes
            on on the simultaneous variations in space and time of these fields, electromagnetism
            constitutes a way of thinking, quite different, quite distinct, from mechanics.        
            ...
    \newline  (Electromagnetism)
        has assimilated without any effort the immense domains of
            optics and radiant heat, unexplained by mechanics, and leads every day to new
            discoveries. Electromagnetism has encompassed most of physics, invaded chemistry
            and grouped together an enormous number of hitherto unrelated facts.
          \end{quotation}
 \textwidth 16cm        
               \par Langevin seems to be unfamiliar with relativistic mechanics, that respect not the
                Galilean but the Lorentz transformation, even though the essential equations, to be
                 discussed in Section 3 below, were already given by Planck in 1906~\cite{PlanckRM},
              and evidently still clings to the 19th Century concept of the `luminifeous aether'
              that is logically incompatible with Einstein's second postulate of the constancy
               of the speed of light for all inertial observers, and that was, in consequence,
              declared to be `superfluous' by Einstein~\cite{Ein1}; this, in spite of
             the fact that, later in the paper, Langevin insists on the fundamental importance
             of the second postulate of special relativity. Langevin appears to be unaware of this
             antinomy in his descriptions of electromagnetism and relativity.
               \par From a modern viewpoint, with knowledge of quantum mechanics and that
                   light consists of photons (massless particles), it can be seen that
                 classical electromagnetism and the wave theory of light are most
                 easily understood
                 as different aspects of the (quantum) mechanics of photons~\cite{JHFEJP,JHFAP}, so that
                     Einstein was correct, from this perspective, to categorise the aether as `superfluous'. 
                \par Langevin mentions the failure of experimental tests to detect motion
                 through the aether without mentioning, by name, the Michelson Morley experiment~\cite{MM}.
                 The incompatibility of the result of this experiment with the aether hypothsis, to which
                 he previously subscribed (and subsequently subscribes) is passed over in silence.
                 Langevin then states the special relativity principle in the following form:
              \begin{quotation}
\textwidth 15cm  
              If various groups of observers are in uniform translational motion
                   with respect to each other (such as observers on the Earth for different 
                  positions of the latter in its orbit) all mechanical and physical phenomena
                   follow the same laws for all the groups of observers. None of them, by experiments
                  inside the material system with which they are correlated, can demonstrate 
                  uniform translational motion of the system.
          \end{quotation}
 \textwidth 16cm 
                 \par Langevin then adds the remark:
              \begin{quotation}
\textwidth 15cm 
                 Concerning electromagnetism it can be stated that the fundamental equations
                 hold for all groups of observers at the same time, that everthing happens for
                 each of them as if they were at rest in the aether.
          \end{quotation}
 \textwidth 16cm
                  The incompatiblity of this statement with the mechanical-aether model of
                 light wave propagation is flagrant but unmentioned by  Langevin!
                 \par Before entering into the detailed discussion of space and time in special
                  relativity Langevin mentions the transformation groups of classical mechanics
                 (Galilean) and of electromagnetism (Lorentz). It is stated,  concerning the
                  latter,
              \begin{quotation}
\textwidth 15cm  
                  {\it This group differs profoundly from the preceding~}(Galilean)~{\it concerning 
                    transformations of space and time.}(italics in the original)
         \end{quotation}
 \textwidth 16cm
              \par Langevin finally remarks, before the passages to be discussed in the following section:
              \begin{quotation}
\textwidth 15cm 
               We must choose; if we wish to maintain the absolute validity of 
                  the equations of classical mechanics, of mechanical processes, as well as the space
                     and time which correspond to them, we must consider as false those of 
              electromagnetism and renounce the admirable synthesis of which I wrote above, 
                 returning in optics, for example, to the emission theory with all the
                 difficulties to which it lead, and which lead to its rejection fifty years ago.
                 If we wish, on the contrary, to retain electromagnetism we must adapt our understanding to
                 the new concepts of space and time, and consider classical mechanics as
                 having no more importance than as a first approximation, largely sufficient
                 moreover, when it concerns a speed that is not more than a few thousand
                 kilometers per second. Electromagnetism, or laws of mechanics respecting
                 the same transformation group, alone allow to go further and occupy
                 the preponderant position which mechanical processes hold in classical mechanics.
         \end{quotation}
 \textwidth 16cm
                    \par Langevin is proposing here relativistic mechanics as some yet-unrealised
                    future project in spite of its then already-known equations~\cite{PlanckRM,Poincare,Mink} and experimental
                   evidence~\cite{Bucherer} for their validity. Some results of such experiments were even
                   shown in the paper by Lorentz in which the transformation named for him
                   by Poincar\'{e} was introduced. The relativistic-kinematical properties of electrons
                  are nowhere mentioned in Langevin's paper. Langevin much exaggerated
                the current antithesis between electromagnetism and classical mechanics,
                resolved, partly by the already-existing laws of relativistic classical mechanics and
                completely by the later introduction of quantum electrodynamics, of which
                classical electrodynamics is only a particular limit. Quantum electrodynamics is, of course,
                although misnamed, for historical reasons, a `quantum field' theory, actually an
                `emission' theory! The essential physical entities are particles (photons and charged particles) not fields.

\SECTION{\bf{Langevin's discussion of simultaneity, causality and space-like and time-like event intervals}}
 An English translation of the passages in Langevin's paper concerning invariant intervals and their related
 space-time geometry follows (italics in the orginal):
              \begin{quotation}
\textwidth 15cm 
  In order to show more clearly the opposition between the two approaches
    \newline (classical mechanics and classical 
    electrodynamics) it is simpler, as proposed by Minkowski, to merge the two distinct ideas
    of space and time into a more general one in describing the world.
   The universe is the totality of all the events it contains. An event is defined
 as what happens (or what exists) at a certain place at a certain time. Given a
 reference system, that is a system of coordinate axes corresponding to a certain 
 group of observers, the position of an event in space and time is given by 
 coordinates referred to this system of axes, three for its spatial position 
 and one for its temporal one.
 Given two events referred to a certain coordinate system, they differ, in
   general, both in space and time, occuring at different places at different 
   instants. To any pair of events there corresponds a spatial distance
   (that  between the points where the events occur) and a time interval.
 `Time' may then be defined by the totality of successive events at the
     same position, for example in the same physical object with its own reference
   system, and `space' may be defined by the totality of simultaneous events.
    This definition of space corresponds to this: The shape of a moving body is 
      defined by the totality of the simultaneous positions of the different
      material elements of which it is constituted, or equivalently, by
    the totality of events which specify the simultaneous presence of the
    different material elements. If it is agreed, with Minkowski, to call the
    {\it world line} of an element of matter, which may be in motion
     relative to a system of reference, the totality of events which succeed
     each other in that element of matter, the shape of a material body, at 
     any instant, is determined by the totality of simultaneous positions
     along the world lines of the diverse material elements that constitute
     the body.
   \par The notion of simultaneity of events which happen at different points is fundamental
        for the very definition of space in the case of an extended body in movement, and this is
      generally true.
     {\it In the normal conception of time, simultaneity is absolute, and considered
      to be independent of the reference system;}~we must analyse in detail the meaning
      of that usually tacit hypothesis. 
      Why do we usually not allow that two events, that are simultaneous for a certain 
       group of observers, may not be so for another group of observers in motion 
        relative to the first, or what is the same thing, why do we not allow that a
      change of reference system might reverse the temporal order of two events?
        \par This comes evidently from our implicit admission that if two events follow
      each other in a certain order, for a given reference system, the first may causally
        affect the conditions under which the second is produced, independently of the
        distance separating them in space. Under such conditions it is absurd to 
       suppose that, for other observers, in another reference system, the second event,
       the effect, could occur before its cause. The absolute character usually
      supposed for the notion of simultaneity stems therefore from the implicit assumption
      that causal influences might propagate at infinite speed, that therefore an event might have
     an instantaneous causal effect at any distance.
     \par The latter hypothesis agrees with mechanical concepts and is required by them,
             because a perfect solid or a non-extensible string connecting the two points 
       where the events occur allows to signal instantaneously the occurence of the first
       event at the position where the second event occurs, 
         and consequently permitting a causal connection between
       the two events. There is therefore consistency between classical mechanics
       and everyday conceptions of space and time in which the simultaneity
        of two spatially-separated events has an absolute meaning.
       \par We are therefore not at all surprised to note that in the transformation
       group that that leaves invariant the equations of classical mechanics
       {\it the time interval between two events is invariant and is measured
            to be the same by all groups of observers, whatever their relative motion.
         \par It is otherwise with spatial intervals:~} it is a simple fact, in 
        accordance with commonsense ideas, that the spatial separation of two events
        does not, in general, have an absolute sense and so depends on the reference 
        system that is considered. 
         \par A simple example shows how the spatial interval between the same
         pair of events may be different for different groups of observers in
         relative motion. Imagine that two objects drop successively through
        a hole in the floor of a cart moving along a road: the two events corresponding
        to the dropping of the two objects through the hole occur at the
        same point for observers in the cart, and at different points for observers
        standing by the road. The spatial separation vanishes for the first
       group of observers, whereas for the second it is equal to the product of 
        the speed of the cart and the time interval between dropping the two
        objects.
        \par It is only in the case that the two events are simultaneous that the spatial
        separation has an absolute sense, and does not change with the choice of cordinate
        system. It follows immediately that the dimensions of an object, for example the
       length of a ruler, have an absolute sense and are the same for observers at rest, or
      in movement, relative to the object. We have noted that effectively, for any observation
      of the ruler it is the distance between two simultaneous positions of the ends of the ruler,
      that is to say of the spatial separation of two simultaneous events, of two simultaneous
       positions of the two ends of the ruler. We saw previously that also simultaneity, like the spatial
       separation of two simultaneous events, has an absolute meaning in the commonsense conceptions
       of space and time.
       \par Given any two successive events, two temporally separated events, a reference system
           can always be found for which the two events are spatially coincident. For an observer
         in this system the two events occur at the same spatial position. Indeed, it is sufficient
       to give these observers a motion relative to the initial system such that, having observed
       locally the first event, they likewise observe locally the second one, the two events occuring
       at the same position relative to them in the two cases. It is sufficient to give the
         group of observers a speed equal to the ratio of the spatial separation of the two
        events to the time interval between the two events in the initial reference
       system, and this is always possible if the time interval between the two events does not
       vanish, if the two events are not simultaneous. 
       \par What may, in this way, be realised in space, the spatial coincidence of two
          events by a suitable choice of coordinate system cannot, as we have seen, be realised for
         temporal intervals because they are absolute and measured in the same manner in all 
         reference systems.
         \par There is an asymmetry in the commonsense concepts of space and time that is removed
            by application of the new concepts. Temporal intervals, like spatial ones, now depend on 
          the reference system from which they are observed.
         \par With the new concepts, only one case remains and must remain where a change of
             coordinate system has no effect. This is the case where two events coincide
          in both space and time. Such a double coincidence must have, indeed, an absolute meaning
          because it corresponds to contiguity of the two events and that contiguity
          may produce a physical phenomenon, a new event, which has necessarily an absolute
          sense. Consider again the previous example. If the two objects which
          leave the cart by the same hole do so simultaneously, if their departures coincide
         both in space and time, there may result a collision which breaks the objects,
         and that collision has an absolute sense in that, in no possible conception of
         the world, electromagnetic or mechanical, such a spacetime coincidence, if it
         exists for one group of observers, could it not exist for another, whatever their motion
         relative to the first. For those that see the cart pass by as well as for those inside
         it, the two objects have broken each other because they passed at the same time
          at the same point. 
       \par With the exception of this special case, it is easy to see that the concepts of
         electromagnetism require a profound re-thinking of the notion of the physical
          universe. The equations of electromagnetism imply, according to their usual 
          formulation, that an electromagnetic disturbance, a light wave, for example,
          propagates with the same speed in all directions, equal to about 
         300,000 km/sec. 
        \par Newly established experimental facts have shown that if the equations are
           exact for a group of observers, they must also be so for any others,
           independently of their motion with respect to the first group.
          There results from this the paradoxical fact that a given light signal
           should move at the same speed relative to different groups of observers in
           motion relative to each other. A first group of observers see a light wave
          propagate in a certain direction with the speed 300,000 km/sec and see 
          another group of observers run after the wave with an arbitary speed and yet,
          for the second group, the light wave moves relative to it at the same
          speed of  300,000 km/sec.
  \par Einstein was the first to show how that necessary consequence of electromagnetic
       theory suffices to determine the characteristics of space and time required by the
        new conception of the universe. One realises, following, what has just been said, that
        the speed of light should play an essential role in the new formulation; it is
       the only velocity which is conserved in passing from one reference system to another,
       and plays in the universe described by electromagnetism the same role as that played by an
       infinite speed in the universe described by mechanics. This will be clearly seen from
       the considerations which follow.
 \par  For any pair of events, change of the reference system modifies 
    the spatial and temporal intervals between them; but to quantify these
     modifications one is led to classify each pair of events into one
     of two large categories for which space and time play symmetric roles.
      The first category contains all pairs of events such that their
     spatial separation is greater than the path followed by light during
    the time interval between the events; that is, such that, if the emission
  of a light signal  accompanies one of the two events, the other one will
   occur {\it before } the passage of the signal coming from the other 
   one. This is an absolute category, that is, if it is true for one frame
   of reference it is true for all.
    The transformation equations required by electromagnetic theory
     show that, in this case, the order of succession of the two events in
     time has no absolute meaning. If, for the first reference system, the two
    events occur with a certain ordering, this ordering may be inverted
    for some observers moving relative to the first with a speed less than
    that of light, that is, with a physically realisable speed.  
  It is evidently impossible that two events, for which the temporal order
   may be inverted in this way, be connected by a relation of cause and
   effect, since if such a relation existed between the two events, certain
   observers would see the cause after the effect, which is absurd.
  Indeed, given that the spatial separation of the two events is greater
   than the path followed by light during an interval equal to their temporal
   separation, the first cannot be associated with the latter unless the causal 
   effect travels at a speed greater than that of light. We should therefore,
    following what was stated previously, eliminate such a possibility: The 
    causal effect, whatever its nature, should not propagate faster than the
    speed of light; there should exist neither messenger nor signal able to
    cover more than 300,000 km per second.
  We conclude from this that an event cannot give rise to
       action-at-a-distance, that its effect can only occur instantaneously
       at the very position at which the event occurs, then later at
      increasing distances, distances increasing at most with the speed 
    of light. The latter therefore plays a role, in this respect, in the
    new conceptual understanding, that is played, according to the
    old understanding, by an infinite speed, representing, in the latter
     case, the maximum speed consistent with causality. 
  \par It is seen from this that the present conflict between mechanics and electromagnetism 
        is only a new manifestation of the opposition of two concepts which followed each other in the
        development of electrical theories: that of the instantaneous action-at-a-distance
        of mechanics and that introduced by Faraday of the property of action through a medium
       by local contact interactions. That old opposition now has reprecussions at the most fundamental 
        level.
       \par Diverse consequences follow from this: first of all it is impossible that a massive object
         moves relative to another with a speed greater than that of light. This paradoxical result
         follows from the velocity addition formula of the new kinematics: the addition of
         any number of velocities less than the speed of light gives always a velocity less
          than that of light. As in the commonsense concept, the addition of an arbitary number of finite 
        velocities yields always a finite velocity.
        \par We can further assert that no action-at-a-distance, for example the force of gravity,
          can propagate faster than light and it is known that this condition is in no way contradicted 
          by presently known astronomical observations.
        \par  Finally, it is necessary to give up the perfect solid of mechanics which gives a method to
         signal instantaneously to a remote location, to establish a causal link propagating faster
         than the speed of light; elastic waves in the most rigid of real solids propagate, in fact,
         with a much lower speed. The important point is that the very concept of a perfect solid, of
         a body all points of which may be set in motion simultaneously, must be rejected.
         \par The previous reasoning may be summarised as follows: if a signal existed that propagated
           with a speed greater than that of light, observers could be found for which the signal could 
         arrive before it was emitted, for which the causal link established by the signal
         would be found to be inverted. We could telegraph into the past, as Einstein said, and we consider
         that would be absurd. 
    The pair of events considered, that have no definite time
     ordering, are therefore necessarily without any possible mutual influence,
      they are truly independent events. It is evident that, being without any causal
       connection, they cannot appear successively in the same material element;
       they cannot belong to the same world line, or to the life of the same
     being. This impossiblity is in agreement with the fact that to be at the
     location of the two events the element of matter must move at a speed 
     greater than that of light. 
     The two events therefore cannot be rendered spatially coincident by any
      choice of reference system, but they may be temporally coincident. Since the order
    of succession may be inverted, there must exist systems of reference for which the 
    two events are simultaneous.
  One may call {\it space-like } the pair of events that have just
     been considered, for which temporal order has no absolute sense, but which are
      spatially separated in an absolute manner. {\it  It is remarkable that, if
      the spatial separation of two events cannot be annuled, it is minimal precisely
       for the reference system relative to which the two events are simultaneous.}
     From which it follows that {\it The spatial
      separation of two events that are simultaneous for a certain group of
      observers is shorter, for them, than for any other group of observers in arbitary
       motion relative to them.}
          This statement contains, as a special case, what has been called
         Lorentz contraction; that is the fact that the same ruler, viewed by different
          groups of observers, some at rest, others in motion with respect to it,
          is shorter for those in motion than for those for which it is at rest.
      We have seen that the length of a ruler, for observers that are in motion 
       relative to it, is defined by the spatial separation of the two positions,
       that are simultaneous for the observers, of the ends of the ruler. That
       distance, according to the above statement, will be shorter for these
       observers than for all others, in particular, those relative to which the
      ruler is at rest.
     It is easily understood also how the Lorentz contraction can be reciprocal,
        that is, how two rulers equal at rest see themselves mutually shortened
         when they slide one against the other, observers at rest relative to one of them
       seeing the other one shortened. This reciprocity holds since the observers at rest
       relative to the two rulers do not define simultaneity in the same way.
       We will find for pairs of events of the second class properties exactly
        correlated with the preceding by permutation of space and time. These events
        that I will call {\it time-like pairs } are defined by the following condition,
         which has an absolute meaning: The spatial separation of two events is less
        than the path followed by light during their temporal separation;
        otherwise said, the second event happens {\it after } the passage of a light
         signal whose emission coincides in space and time with the first event.
          This introduces a time asymmetry between the two events; the first precedes
         the passage of the light signal emitted in space-time coincidence with the 
          second event, while the second follows the passage of the signal whose emission
           accompanies the first. A causal connection may exist, at least by a light signal,
           between the two events. The second could have been informed of the first, and this
            requires that the order of succession has an absolute meaning, not to be inverted
           by any change of reference system. It is seen immediately that such an
           inversion requires a speed greater than that of light for the second
            system of reference relative to the first.
            Two events between which exists, in this way, a real possiblity of
          of causal connection, if they may not be rendered coincident in time, can can always
          be be rendered coincident in space by a suitable choice of reference system.
          In particular, if two events belonging to the same world line have
         an absolute order in the history of an element of matter, they coincide in 
         space for all observers at rest relative to that element.
 { \it  Consistent with what is stated above, if the temporal interval
   of two events cannot be annuled, it is minimal precisely for the reference
        system relative to which the two events are in spatial coincidence.} 
        From which it follows:
 { \it   The temporal interval between two events which coincide in space, and which
        occur successively at the same position for a certain reference system, is less, for this
        reference system, than for any other in uniform translational motion relative to the
         first.}
           \end{quotation}
 \textwidth 16cm 
  \par In the passages quoted above Langevin contrasts the space-time
            concepts of `mechanics' following from the Galilean space and
          time transformation equations (GT):
 \ba
   \rx' & = &\rx-v\rt \\
   \rt' & = & \rt \\
   \ry' & = &  \ry,~~~~~\rz' = \rx
 \ea
  with those of the `transformation equations of electromagnetic
          theory' i.e. those following from, what will be termed in this paper, the `generic'
      space-time Lorentz transformation (LT) as originally derived by Larmor~\cite{Larmor},
      Lorentz~\cite{Lorentz} and Einstein~\cite{Ein1} :
 \ba
   \rx' & = & \gamma [\rx-v\rt] \\
   \rt' & = & \gamma [\rt- \frac{v \rx}{c^2}] \\
   \ry' & = &  \ry,~~~~~\rz' = \rx
 \ea
     where $\gamma \equiv 1/\sqrt{1-\beta^2}$; $\beta \equiv v/c$ and
          $c$ is the vacuum speed of light. In the transformation equations (2.1)-(2.6), the coordinates
         of an event in the inertial frame S are ($\rx$,$\ry$,$\rz$,$\rt$) and in the
        inertial frame S' are ($\rx'$,$\ry'$,$\rz'$,$\rt'$). The spatial axes are parallel and S' moves
        with speed $v$, relative to S, in the positive $\rx$-direction.
      \par After a general discussion of time intervals, space
          intervals and simultaneity as well as causality concepts based
          on Einstein's postulate of the constancy of the speed of
          light and the relativistic parallel velocity addition
          formula, space-like and time-like event intervals are
          introduced. Qualitative consideration of these intervals
          leads to the introduction of the new space-time  `effects'
          of special relativity: relativity of simultaneity (RS),
          length contraction (LC) and time dilatation (TD). In the
          following, it is first shown how the space-time geometrical effects discussed by Langevin in the 
     above passages may be derived from the generic LT (2.4)-(2.6). The former are then compared
     and contrasted with predictions for observations of synchronised clocks at rest in the frames S or S'.
    In the light of the results of these calculations, Langevin's
          comparison  of the `mechanical' and `electromagnetic'
          concepts of space and time are re-examined, in particular his erroneous
          conflation of concepts of space-time geometry and physical causality. 
 \begin{figure}[htbp]
\begin{center}\hspace*{-0.5cm}\mbox{
\epsfysize15.0cm\epsffile{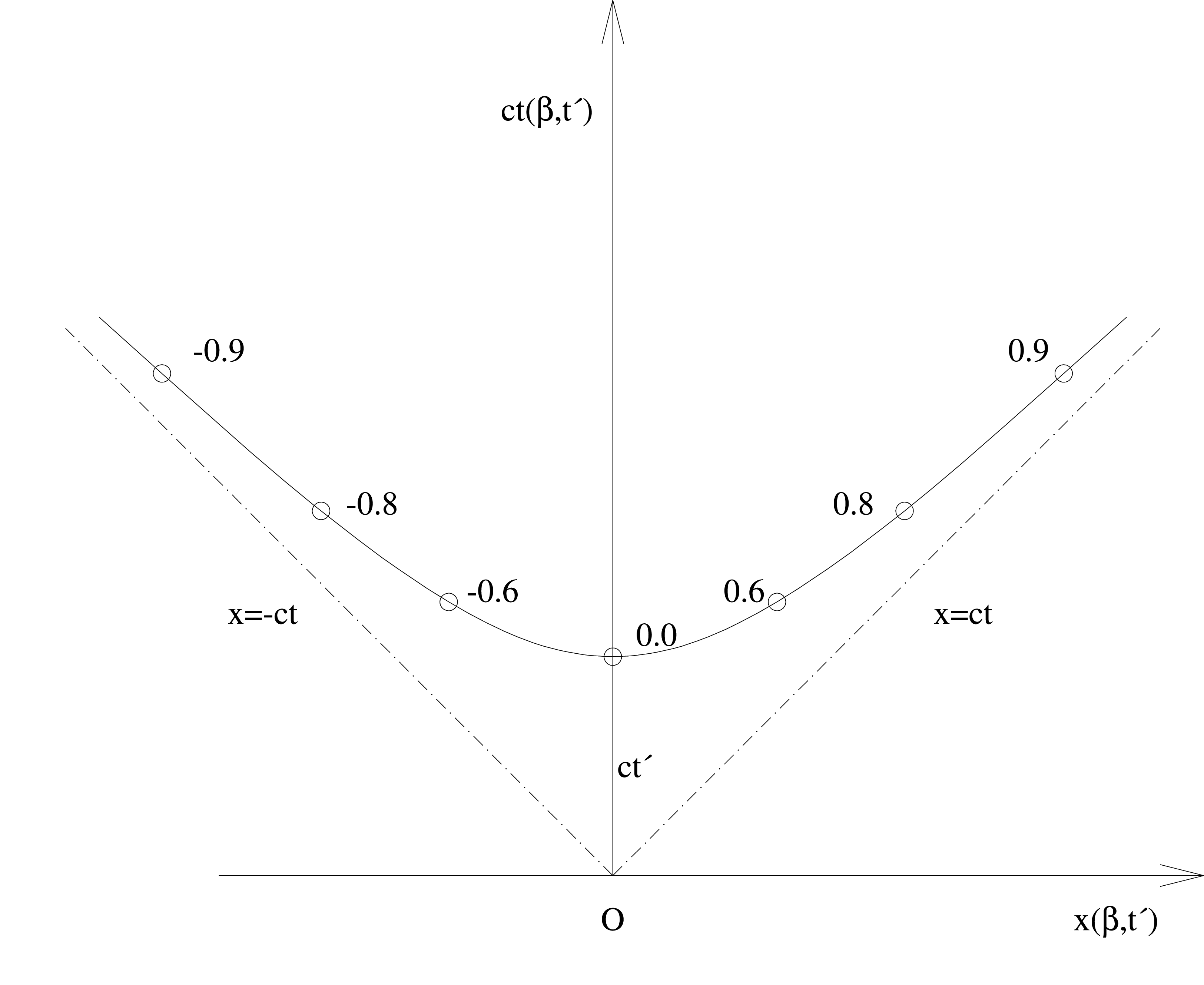}}
\caption{ {\em Hyperbola (2.12) giving the locus, for different values of $\beta$, of the world line
 in S of an event ($0$,$t'$) at the origin of S'. The light-cone projections in the $ct$ versus $x$
 plane: $x = \pm ct$, which are the asymptotes of the hyperbola, are indicated as dashed lines.
  Points on the hyperbola marked with small circles are labelled with the corresponding values
   of $\beta$. The value of ct is minimum when $\beta = 0$ (TD).}}
\label{fig-fig1}
\end{center}
\end{figure} 
    As in Ref.~\cite{JHFSTP3} a roman font is used for the space and time coordinates in the
    generic LT to distinguish them from symbols representing the spatial positions and recorded
    times (epochs)\footnote{An `epoch' is a number, derived from a clock, that is 
       in one-to-one correspondence with a certain instant of time in a given reference frame.
         The operational definition of a physical time interval is then difference of two
         epochs defined in this way.}
      of the spatially-separated and and synchronised clocks to be discussed below.
     Throughout this paper, only events
      with coordinates: ($\rx$,$0$,$0$,$\rt$) $\equiv$ ($\rx$,$\rt$),
    ($\rx'$,$0$,$0$,$\rt'$) $\equiv$ ($\rx'$,$\rt')$ in the inertial frames S and S' respectively,
      lying on the common $\rx$-$\rx'$ axis of
    these frames are considered. \
    \par As first pointed out by Poincar\'{e}~\cite{Poincare} and Minkowski~\cite{Mink}, the generic LT allows, in virtue of the
      identity $\gamma^2-\gamma^2\beta^2 \equiv 1$ to demonstrate the existence on invariant intervals
      $\Delta \rx$, $\Delta \tau$ defined as:
    \beq
   (\Delta \rs)^2 \equiv -c^2(\Delta \tau)^2  \equiv (\Delta \rx)^2-c^2 (\Delta \rt)^2
     = (\Delta \rx')^2-c^2 (\Delta \rt')^2 
     \eeq
    where $\Delta \rx \equiv \rx_2-  \rx_1$  {\it etc} and ($\rx_1$,$\rt_1$),($\rx_2$,$\rt_2$),($\rx'_1$,$\rt'_1$)
    and ($\rx'_2$,$\rt'_2$) are  the coordinates of arbitary events in S and S'
     connected by the
     transformations (2.4) and (2.5).
    \par If $\Delta \rx > c \Delta \rt $, Event 1 and Event 2 are causally disconnected, in the sense
    that an observer moving at a speed less than that of light cannot be present at both events.
     Also only light signals sent from the position of Event 1 for epochs
     $\rt_1-\Delta \rx/c+\Delta \rt < \rt_1$
    can arrive at the position of Event 2 before the epoch $\rt_2$. In this case $\Delta \rs$ is real ,
     $\Delta \tau$ is imaginary and the two events are said to have a space-like separation (Langevin's
     `space-like pair').
    In the contrary case, $\Delta \rx < c \Delta \rt $ an observer moving at a speed less than that
    of light can be present at both events and a light signal emitted
    at the position of Event~1
    at the epoch $\rt_1-\Delta \rx/c+\Delta \rt > \rt_1$ arrives at the position of Event 2 at the epoch
    $\rt_2$. The interval  $\Delta \rs$ is now imaginary and  $\Delta \tau$ is real and the
     events have a time-like separation (Langevin's `time-like pair'). It is clear from Eq. (2.7)
  that the classification of
     a pair of events as space-like or time-like separated is valid in all inertial frames
     with coordinates connected by (2.4) and (2.5), since the parameter $v$ does not appear in (2.7).
     \par It is convenient to now consider intervals containing the event ($\rx'$,$\rt'$) = ($0$,$0$)
      situated at the origin of coordinates in S'. This is one of the sequence of events
 ($0$,$\rt'$) specifying the world line of a ponderable physical object placed at the origin 
      of S'. The coordinates of the world line of such an object in S, ($\rx(\beta,\rt')$, $\rt(\beta,\rt')$)
      as given by (2.4) and (2.5) are:
      \ba
     \rx' & =  &  \gamma[\rx(\beta,\rt')- v \rt(\beta,\rt')] = 0 \\
   \rt' & = & \gamma[\rt(\beta,\rt')-\frac{v \rx(\beta,\rt')}{c^2}]
      \ea
     Combining (2.8) and (2.9) gives the TD relation
   \beq
    \rt(\beta,\rt') = \gamma \rt'
   \eeq
    so that
    \beq
    \rx(\beta,\rt') = \gamma \beta c \rt'
   \eeq   
     Since  $\gamma^2-\gamma^2\beta^2 \equiv 1$ (2.10) and (2.11) yield
 the time-like invariant interval relation
      \beq
     c^2 (\rt')^2 = c^2 \rt(\beta,\rt')^2- \rx(\beta,\rt')^2 
     \eeq
     For a fixed value of $\rt'$, (2.12) gives a hyperbola on the plot of $c\rt(\beta,\rt')$ versus
     $ \rx(\beta,\rt')$ as shown in Fig. 1.
     According to Eq. (2.10) and (2.11), as stated by Langevin, the minimum value of the time
     interval $\rt(\beta,\rt')$ occurs, for any value of $\rt'$, at
    $\beta = \rx = 0$ and $\rt' = \rt$.
 \begin{figure}[htbp]
\begin{center}\hspace*{-0.5cm}\mbox{
\epsfysize15.0cm\epsffile{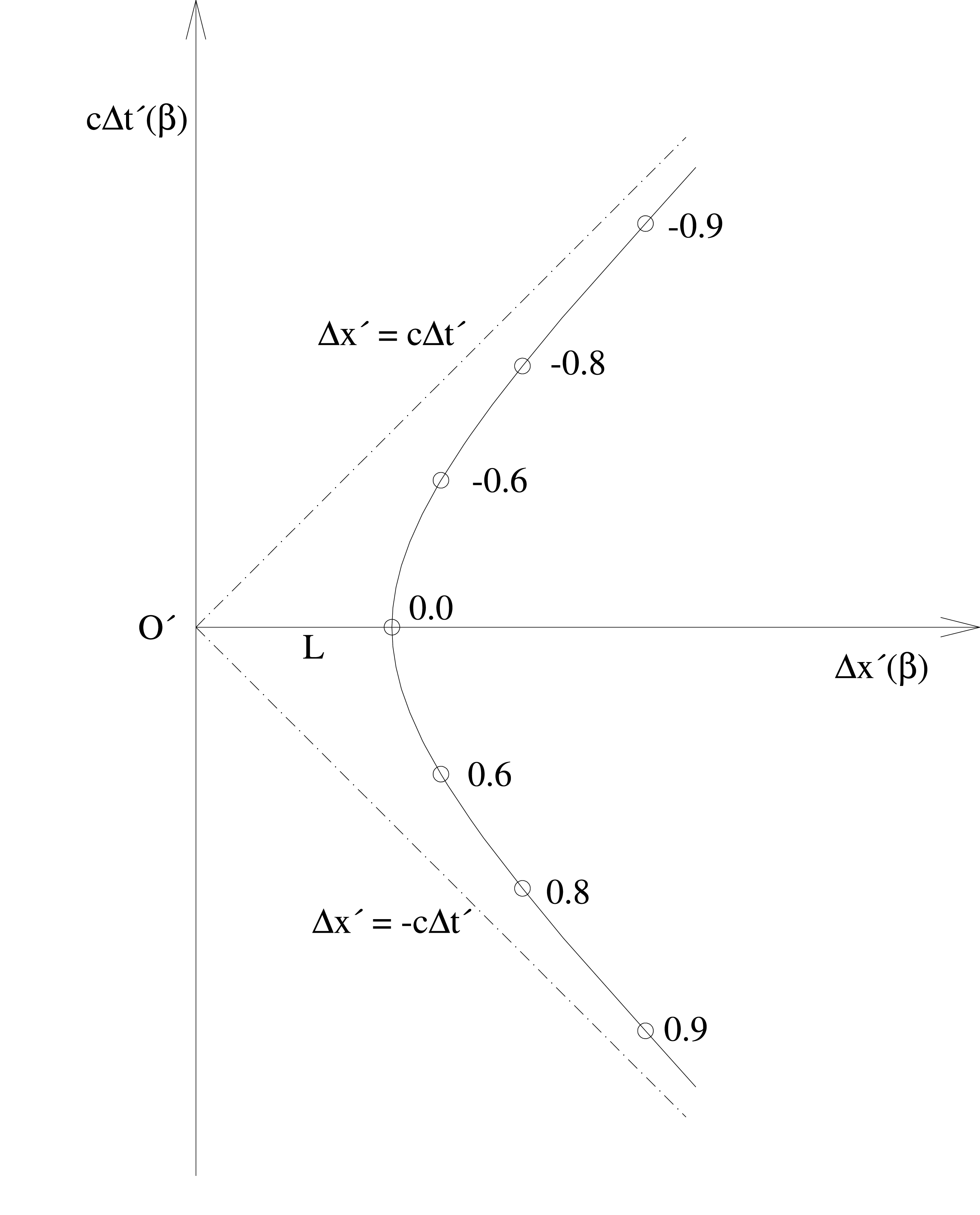}}
\caption{ {\em Hyperbola relating, according the Eq. (2.19), for various values of $\beta$, $\Delta x'$ and
  $c\Delta t'$, the spatial and temporal intervals in the frame S' corresponding to the simultaneous 
    events ($0$,$t$) and ($L$,$t$) in the frame S. Points on the 
  hyperbola marked with small circles are labelled with the corresponding values
   of $\beta$. The sign of $\Delta t'$ depends the value of $\beta$ (RS) and the
    minimum value of  $\Delta x'$ occurs when $\beta = 0$ (LC).}}
\label{fig-fig2}
\end{center}
\end{figure} 
     \par Consider now the interval between two simultaneous and space-like separated events with 
      coordinates in
    the frame S: ($\rx_1$,$\rt_1$) = ($0$,$\rt$), ($\rx_2$,$\rt_2$) = ($L$,$\rt$). Application
    of (2.4) and (2.5) gives:
     \ba
     \rx'_1(\beta,\rt) & = & -\gamma v \rt \\
    \rt'_1(\beta,\rt) & = & \gamma \rt \\
 \rx'_2(\beta,\rt) & = & \gamma[L- v\rt] \\
   \rt'_2(\beta,\rt) & = & \gamma[\rt- \frac{vL}{c^2}]
   \ea
     Combining (2.13) and (2.15) and (2.14) and (2.16) gives:
    \ba
     \Delta  \rx'(\beta) & \equiv & \rx'_2(\beta,\rt) -  \rx'_1(\beta,\rt) = \gamma L \\
 \Delta  \rt'(\beta) & \equiv & \rt'_2(\beta,\rt) -  \rt'_1(\beta,\rt) = -\frac{\gamma \beta L}{c}
        = -\frac{\beta \Delta  \rx'(\beta)}{c}
     \ea
  which yield the space-like invariant interval relation
  \beq
    L^2  = \Delta \rx'(\beta)^2 -c^2 \Delta \rt'(\beta)^2  
  \eeq
   Eq. (2.17) displays the relativistic LC effect, while, according to (2.18), the
   Events~1 and 2, by construction simultaneous in S, are not so in S' if $\beta > 0$. This
    is the RS effect. The hyperbola (2.19) in the $c\Delta  \rt'$ versus
     $ \Delta  \rx'$ plot is shown in Fig. 2. As remarked by Langevin,
    the minimum spatial separation, $L$,
      of the events~1 and 2 in S' occurs, according to (2.17) and (2.18),
     when $v =  \Delta  \rt'(0) =  \Delta \rt =  0$, so that the events
    are simultaneous and the frames S and S' are the same. For any
    value of $v$ greater than zero, the Events~1
    and Event~2 are not simultaneous and their spatial separation is greater in S' than in S. 
       \par The qualitative effects, concerning time-like and
    space-like separated events, discussed by Langevin, in the passages quoted above, have now been derived
     by application of the generic LT (2.4) and (2.5). In particular, the TD, LC and RS effects have all been 
     obtained as rigorous results of the application of certain mathematical operations on the equations
     of the generic LT. As will be further discused below, TD is the result of the projection $\rx' = 0$,
     while LC and the associated RS effect are given by the $\rt = 0$ projection. What however is 
     essential for the physics, and what is not taken into account in these procedures, is the operational
     meaning of the equations, that is, the exact correspondence between the time symbols in the equations
     and the epochs of actual physical and synchronised clocks~\cite{JHFSTP1}. Synchronisation
     is a perfectly-controlled and purely mechanical procedure~\cite{JHFCRCS} and the space-time 
     transformation equations must be written in such a way that synchronised clocks are correctly
     described by them. As will be seen, in order to achieve this, certain constants must be 
      added to the right sides of (2.4) and (2.5) if a synchronised clock, not placed at the origin
       of coordinates in S', is considered. Interestingly enough, as will be further discussed below,
       Einstein clearly stated the necessity to include such constants in the original 1905 special
      relativity paper~\cite{Ein1}!
    \par The LT will now be used to predict observations
 of the epochs of two spatially separated and synchronised clocks C'$_1$ and  C'$_2$ at rest in S'at the
      positions\footnote{A mathmode font is now used for the space-time coordinates}
       $x'_1 = 0$, $x'_1 = L'$ respectively. Using the same spatial coordinate system as in 
       generic LT for both clocks, the equations of motion of the clocks in S are:
      \ba
         x_1(\beta,t') & = & v t_1(\beta,t')   \\
   x_2(\beta,t') & = & v t_2(\beta,t')+L 
    \ea
    where $t'$ is the common epoch of C'$_1$ and  C'$_2$ and $L \equiv  x_2(t_2 =0)$ is a constant,
    independent of both $\beta$ and $t'$. In (2.20) and (2.21) $ t_1(\beta,t')$ and   $t_2(\beta,t')$
    are the epochs of C$_1$ and C$_2$, which are clocks similar to C'$_1$ and  C'$_2$ but at rest
    at arbitary positions in the frame S.
    \par For the clock C'$_1$ situated at the origin of S', the appropriate LT
      is the same as (2.8) and (2.9) above: 
     \ba
     x'_1 & =  &  \gamma[x_1(\beta,t')- v t_1(\beta, t')] = 0 \\
   t' & = & \gamma[t_1(\beta,\rt')-\frac{v x_1(\beta,t')}{c^2}]
      \ea
     from which may be derived the relations similar to (2.10) and (2.11):
   \ba
    t_1(\beta,t') & = & \gamma t'  \\
    x_1(\beta,t') & = & \gamma \beta c t'
    \ea
     Since  C'$_2$ has the fixed position $x'_2 = L'$ for all values of $t'$ and $v$, while
     its space-time coordinates in S obey the equation of motion (2.21), the space
     transformation equation consistent with (2.22) in the limit $L = L' =0$ (and therefore
     using the same spatial cordinate systems) is:
     \beq
         x'_2-L'  =   \gamma[x_2(\beta,t')-L- v t_1(\beta, t')] = 0
    \eeq
      while the corresponding time transformation equation is:
     \beq
   t'  =  \gamma \left[t_2(\beta,\rt')-\frac{v[x_2(\beta,t')-L]}{c^2}\right]
  \eeq
     These transformation equations give parametric equations of a
    hyperbola
    analogous to (2.24) and (2.25):
   \ba
    t_2(\beta,t') & = & \gamma t'  \\
    x_2(\beta,t') & = & \gamma \beta c t'+L 
    \ea
    It follows from (2.24) and (2.28) that:
     \beq
    t_1(\beta,t') = t_2(\beta,t') \equiv   t(\beta,t') = \gamma t' 
    \eeq
    Thus the clocks C'$_1$, C'$_2$, C$_1$ and C$_2$ which, according to the LTs
     (2.22),(2.23),(2.26) and (2.27), are mutually synchronised so that $t'= t_1(\beta,0)=
     t_2(\beta,0) = 0$ when $x_1 = x_2-L = 0$, remain so at all later epochs and for all
    values of $v$ ---there is
   no RS as in the calculation above that leads to Eq.(2.18).
   The equations (2.24),(2.25),(2.28) and (2.29) for the space-time
  coordinates in the frame S and the TD relations Eq.~(2.30) yield the time-like invariant interval relations:
    \beq 
     x_1(\beta,t')^2-c^2t(\beta,t')^2 = [ x_2(\beta,t')-L]^2- c^2t(\beta,t')^2 = c^2 (t')^2
     \eeq
   These hyperbolae, which represent the loci of points on the world lines of the
   clocks C'$_1$ and C'$_2$ in S for a fixed value of $t'$ and various values of
    $\beta$ are plotted in Fig. 3. 
    \par Eqs.~(2.25) and (2.29) give 
      \beq 
     x_2(\beta,t')- x_1(\beta,t')= L   
     \eeq
      demonstrating the velocity independence of the spatial separation of
    C'$_1$ and C'$_2$ in S. This invariance is obvious on inspection of Fig. 3, on
    comparing events with the same value of $\beta$ on the hyperbolae
    corresponding to C'$_1$ and C'$_2$. 
    The special case $\beta = 0$ in 
   (2.32) gives:
     \beq 
     x_2(0,t')- x_1(0,t') = x'_2-x'_1 \equiv L'= L   
     \eeq
 \begin{figure}[htbp]
\begin{center}\hspace*{-0.5cm}\mbox{
\epsfysize15.0cm\epsffile{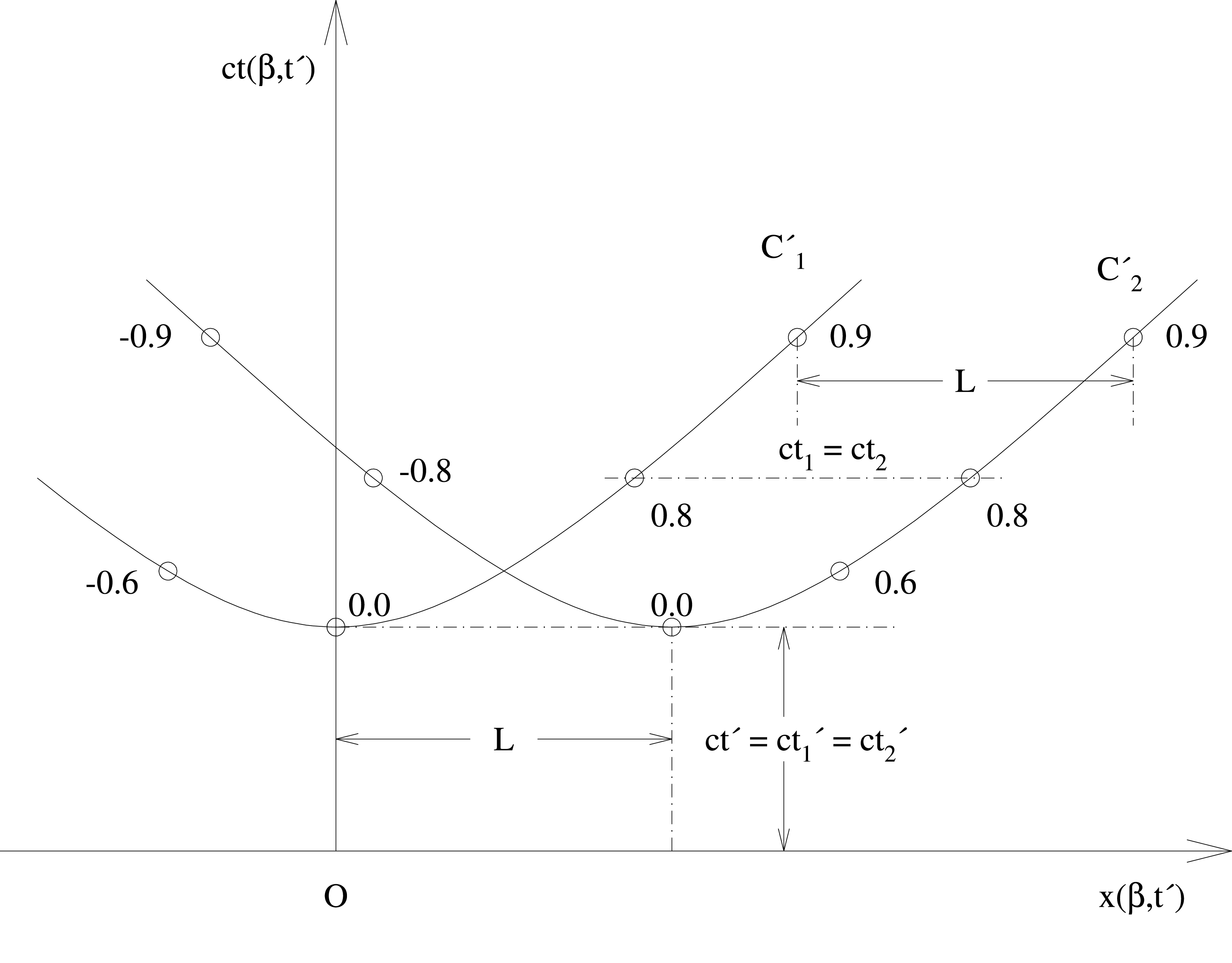}}
\caption{ {\em Hyperbolae in Eq.~(2.31) giving the loci, for different values of $\beta$, of the world lines in S
 of two synchronised clocks, C'$_1$ (at $x'= 0$) and  C'$_2$ (at $x'= L$) with a common epoch $t'$.
  Points on the 
  hyperbolae marked with small circles are labelled with the corresponding values
   of $\beta$. The absence of any RS effect ($t_1 = t_2$ for any value of $\beta$) or LC effect
    ($x_1-x_2 = L$ for any value of $\beta$) is evident from inspection of this figure.}}
\label{fig-fig3}
\end{center}
\end{figure} 
 \begin{figure}[htbp]
\begin{center}\hspace*{-0.5cm}\mbox{
\epsfysize15.0cm\epsffile{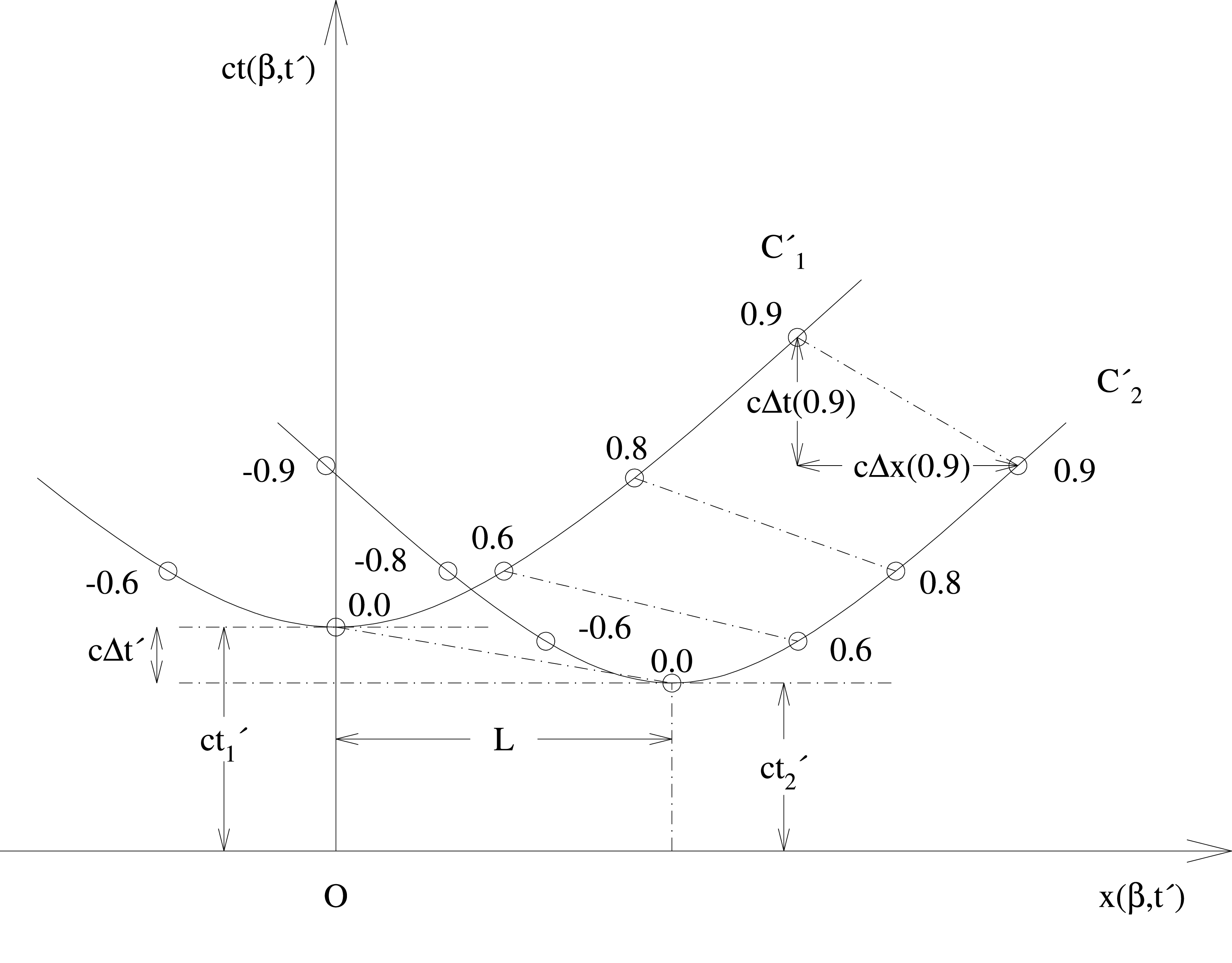}}
\caption{ {\em As Fig. 3 but with non-simultaneous events ($t'_1 > t'_2$) on the world lines
    of C'$_1$ and  C'$_2$. $\Delta t = t_1-t_2$ is non-vanishing and positive for all values of
  $\beta$ and is an increasing function of $|\beta|$ (no RS).}}
\label{fig-fig4}
\end{center}
\end{figure} 
   The hyperbolae giving the loci of world-line points of C'$_1$ and C'$_2$ in S
       for fixed values of the epochs $t'_1$ and $t'_2$ of C'$_1$ and C'$_2$  ($t'_1 > t'_2$)
      are plotted in Fig. 4. For  $t'_1 \ne$ $t'_2$ as in this figure, the corresponding
      equations analogous to (2.24),(2.25), (2.28) and (2.29) give
     \ba 
       \Delta t & \equiv & t_1-t_2 = \gamma \Delta t' \equiv \gamma (t'_1-t'_2) \\ 
      \Delta x & \equiv & x_2-x_1 = L -\gamma \beta c \Delta t'
     \ea
      These equations show that the signs of $ \Delta t$ and $ \Delta t'$ are the same for all
     values of $\beta$ (no RS) and that for given values of $\Delta t'$ and $L$, $\gamma \beta$
      may be chosen so $\Delta x$ takes any value:
       \beq
      \gamma \beta(\Delta x) = \frac{L -\Delta x}{c \Delta t'}  
       \eeq
      This equation requires that $L > \Delta x > 0$ for $0 <\gamma \beta < L/(c\Delta t')$
    and  $0 > \Delta x > -\infty$
      for $ L/(c\Delta t') < \gamma \beta < \infty$ while  $\Delta x > L$ corresponds to
     negative values of  $\gamma \beta$.  
       Comparing, in Fig. 4, the values of $t_1$ and $t_2$ for different
      values of $\beta$, it can be seen that contrary to the predictions of Eq. (2.19)
      and Fig. 2, there is no value of $\beta$ for which the non-vanishing value
      of $\Delta t'$ shown corresponds to a vanishing value of
       $\Delta t$,  
        and that $t_1 > t_2$ (no change in the time ordering of the events,
       even though they are space-like separated) for all values of $\beta$. This is in 
       direct contradiction to Langevin's statement
              \begin{quotation}
\textwidth 15cm 
      The transformation equations required by electromagnetic theory
       show that, in this case, the order of succession of the two events in
        time has no absolute meaning.
              \end{quotation}
\textwidth 16cm 
  and the prediction of (2.18), which shows that
        $\Delta \rt'\ne 0$ when $\Delta \rt = 0$ for all non-zero values of $\beta$, and that, as seen in 
     Fig. 2, $\Delta \rt'$ changes sign when $\beta$ changes sign.
      \par There is a clear incompatibilty between the predictions of (2.31)-(2.33) 
     and those (2.17)-(2.19) resulting from application of the generic LT to transform
     the space-time coordinates of the clock C'$_2$. Note that in deriving (2.31)-(2.33)
     the same LT is applied for the clock C'$_1$ as in deriving  (2.17)-(2.19).
      According to Eq. (2.17), the spatial interval in S' between the S frame events
     Event~1: ($0$,$\rt$) and  Event~2: ($L$,$\rt$) is greater than $L$ (LC) and according to (2.18)
      the temporal interval between the two events, which vanishes in the frame
      S, does not do so in the frame S' for any $\beta > 0$.
      Clearly Events~1 and 2 in S are not associated via Eqs. (2.17)-(2.19) with space-like separated events
      on the world lines of the synchronised clocks C'$_1$ and C'$_2$ for which, by hypothesis,
       $\Delta t' = 0$. This incompatibility arises because the LT of Eqs.(2.15) and (2.16)
       corresponding to the subsitution $\rx \rightarrow L$ in the generic LT (2.4) and (2.5)
       {\it manifestly does not describe a synchronised clock at the location  $\rx = L$},
     as do Eqs. (2.26) and (2.27) which give, for C'$_2$: $t_2 = t' = 0$ when $x_2 = L$
      and for C'$_1$: $t_1 = t' = 0$ when  $x_1 = 0$. Indeed, elementary consideration
     of the very definition of an inertial frame, shows that Eqs. (2.17)-(2.19)
      cannot correctly describe events on the world lines of such clocks. The clocks
      C'$_1$ and C'$_2$ are at rest in the frame S', separated by the distance $L'$:
       $x'_1 =0$,  $x'_2 =L'$, independently of both velocity and time. Their
     equations of motion in the frame S are, respectively,
     $x_1 = vt_1$ and $x_2 = vt_2+L$ where $L \equiv x_2(t_2=0)$ is a constant independent
    of both velocity and time.  According to Eq. (2.17), $L'$ changes as a function 
     of $\beta$ for a given value of $L$, where the latter parameter is independent
     of $\beta$,  contradicting the initial hypothesis that
     C'$_1$ and C'$_2$ have fixed positions in S'. In fact, as discussed in
    Refs.~\cite{JHFLLT,JHFSTP1,JHFCRCS,JHFACOORD,JHFUMC} the LT of Eqs. (2.15) and (2.16)
      corresponds to clock that is,{ \it by construction, desynchronised from}  C'$_1$
      {\it by the amount} $-\gamma \beta L/c$. It is this built-in offset of the epoch
     of  C'$_2$ that is erroneously identified as a physical RS effect.
      \par It is instructive to consider the quantity $(\Delta\rx')^2-c^2(\Delta \rt')^2$ appearing in
       Eq. (2.19), in the case that the intervals are those between events on the world lines
       of the synchronised clocks C'$_1$ and C'$_2$ introduced
     above. For this, Eqs. (2.4) and (2.5) with $\rx_1 = 0$,  $\rx_2 =
    L$, as used to give Eqs. (2.13)-(2.16) are replaced by the
    equations (2.22),(2.23) for C'$_1$ and  (2.26) with $L = L'$ and (2.27) for C'$_2$ :
        \ba
    x'_1(\beta,t_1) & =  &  \gamma[x_1- v t_1]  \\
   t'_1(\beta,t_1) & = & \gamma[t_1-\frac{v x_1}{c^2}] \\
    x'_2(\beta,t_2) & =  &  \gamma[x_2-L+ v t_2] +L \\
   t'_2(\beta,t_2) & = & \gamma[t_2-\frac{v(x_2-L)}{c^2}]
      \ea
    These equations give the spatial and temporal intervals in S':
    \ba
 \Delta x'(\beta,t_2,t_1) & = & \gamma(\Delta x-L)-\gamma v \Delta t+L \\
 \Delta t'(\beta,t_2,t_1) & = & \gamma\Delta t-\frac{\gamma v}{c^2}(\Delta x-L)
    \ea
    where $\Delta x \equiv x_2 - x_1$, $\Delta t \equiv t_2 - t_1$ and
    \ba
       \Delta x'(\beta,t_2,t_1) & \equiv &  x'_2(\beta,t_2)- x'_1(\beta,t_1) \\
 \Delta t'(\beta,t_2,t_1) & \equiv &  t'_2(\beta,t_2)- t'_1(\beta,t_1)
   \ea
     (2.41) and (2.42) then give, instead of (2.19):
    \ba
    \Delta \tilde{s}(\beta,t_2,t_1)^2 & \equiv & \Delta x'(\beta,t_2,t_1)^2 - c^2 \Delta t'(\beta,t_2,t_1)^2
       \nonumber \\ 
       &  = & (\Delta x-L)^2+2 \gamma(\Delta x-L)L+L^2-2\gamma v L \Delta t-c^2(\Delta t)^2 
    \ea
      Note that the velocity-dependent interval $ \Delta \tilde{s}$ is not a Lorentz invariant.
       Setting $t_1=t_2 \equiv t$, $\Delta t = 0$ and $\Delta x = L$  for comparison with (2.13)-(2.19)
      gives:
     \beq
      \Delta \tilde{s}(\beta,t)^2 \equiv \Delta x'(\beta,t)^2 - c^2 \Delta t'(\beta,t)^2
      = L^2
     \eeq
     where $\Delta x'(\beta,t,t) \equiv \Delta x'(\beta,t)$ etc. 
      Or, since from (2.42) when $\Delta t = 0$ and $\Delta x = L$ then $\Delta t'(\beta,t) = 0$, 
   \beq
    \Delta x'(\beta,t) = L
  \eeq
     to be contrasted with the LC prediction of Eq. (2.17) that $\Delta \rx'(\beta,t)$ is velocity dependent.
      Notice that, according to (2.47), and unlike (2.17), $\Delta x'(\beta,t)$ is independent of 
      both $\beta$ and $t$, as required by the assumed initial
     conditions of the problem considered.
      As previously remarked, the
     velocity dependence of  $\Delta x'$ in (2.17) means that it is impossible that the events~1 and 2
      considered in (2.13)-(2.19)  can lie on the world lines of objects at fixed positions
    in S', like the clocks C'$_1$ and C'$_2$.
    \par One may now assess the relevance of Langevin's discussion of space-time geometry, in the
    passages quoted above, to the description of different observations of
     the epochs of spatially-separated and synchronised
    clocks at rest in the same inertial frame. Briefly stated, all of the properties of time-like
    intervals stated by Langevin describe correctly the observed behaviour of a single clock in
    uniform motion, ---in particular the experimentally-well-verified TD effect--- whereas all 
    the discussion of pairs of space-like separated events, leading to the prediction of RS and LC,
    is erroneous. This is because the generic LT (2.4),(2.5) does not correctly describe 
    a synchronised clock at rest in S' at any postion other than that of the origin of
    coordinates $\rx'= 0$. The LT (2.26),(2.27) that does correctly describe a synchronised
    clock at $x'= L$, differs from the generic LT (2.4) and (2.5) by the inclusion of
    additive constants on the right sides of the equations:
     \ba
   x' & = & \gamma [x-vt] +X(\beta,L)\\
   t' & = & \gamma [t- \frac{v x}{c^2}] +T(\beta,L)
 \ea     
  where
  \beq
   X(\beta,L) \equiv (1-\gamma)L,~~~~  T(\beta,L) \equiv \frac{\gamma v L}{c^2}
  \eeq
   The necessity to add such constants to the right sides of the generic LT in order
   to correctly describe synchronised clocks at different spatial positions was clearly
   stated by Einstein in the 1905 Special Relativity paper, immediately
    after his derivation of the LT in the passage\footnote{See Section X of Ref.~\cite{Ein1}.}:
              \begin{quotation}
\textwidth 15cm 
      \par `Macht man \"{u}ber die Anfangslage des bewegten Systems und \"{u}ber
            den Nullpunkt von $\tau$ keinerlei Voraussetzung, so ist auf den
                rechten Seiten dieser Gleichungen je eine additive Konstante zuzuf\"{u}gen.'
       \newline
     \par `If no assumption is made concerning the initial position of the moving system
            and as to the zero point of $\tau$, an additive constant is to be placed
       on the right side of each of these equations.'
\textwidth 16cm
       \end{quotation}

      \par The variable $\tau$ is  $\rt'$ in the notation of the present paper, and `these equations'
      are equivalent to the generic LT (2.4)-(2.6) above.
    To the present author's best knowledge, this admonition was respected
     neither by Einstein himself, nor any other author, before his own
     work in Refs.~\cite{JHFLLT,JHFSTP1,JHFSTP2,JHFCRCS,JHFACOORD,JHFUMC}.
    \par The RS and LC effects are  rigorously correct mathematical consequences of certain projection
     operations performed on the generic LT (2.4) and (2.5). However the additional constants in
     Eqs.(2.48) and (2.49) needed to correctly describe a synchronised clock at $x' = L$ change completely
     the physical consquences of the projection operations when space-like separated events are involved.
      An example of this is the $\Delta t =0$ projection corresponding to a length measurement discussed
      above where the generic LT gives the LC relation (2.17) while the correct LT (2.26),(2.27)
      for C'$_2$ gives instead (2.47) showing no LC effect.
      \par Intoducing unidimensional [L] space-time coordinates by multiplying times by the speed
       of light: $\rx_0 \equiv c\rt$,: $\rx'_0 \equiv c\rt'$ enables the generic LT to be written in a 
     manifestly space-time symmetric manner\footnote{The pair of equations (2.51) and (2.52) is invariant
      under the space-time exchange operations $\rx_0 \leftrightarrow \rx$, $\rx'_0 \leftrightarrow \rx'$.
       Indeed, as shown in Ref.~\cite{JHFAJP2}, requiring invariance under these operations
       is a sufficient condition to derive the LT once space-time homogeneity or
       single-valuedness of the equations is also required.}
       \ba
        \rx' & = & \gamma(\rx-\beta\rx_0) \\
   \rx'_0 & = & \gamma(\rx_0-\beta\rx)
       \ea
       As discussed in Ref.~\cite{JHFAJP1} the various space-time `effects' of special relativity
       are related to the projective geometry of these equations, in particular, the four possible
       projections:
     \[\rx = 0,~~~\rx' = 0~~~\rx_0 = 0~~~\rx'_0 = 0 \]
       As summarised in Table 1, each projection gives a pair of parametric equations
       specifying a hyperbola in space-time corresponding to either a space-like or time-like
       invariant interval relation. However, because of the effect of the additive constants
       in Eqs. (2.48) and (2.49) only the projections $\rx = 0$ and $\rx' = 0$ describe correctly
      the observed behaviour of spatially-separated synchronised clocks. The space-time `effects'
      corresponding to the four projections are\footnote{In Ref.~\cite{JHFAJP1} the acronym
      LFC for `Lorentz-Fitzgerald Contraction' was used instead of LC, and SD for `Space Dilatation'
      instead of LD.}
      \par 
      ~~~~~~~~~~~~~~~~~~~~~~~~~~~$\rx' = 0$:~~~~~~Time Dilatation (TD)
      \par
      ~~~~~~~~~~~~~~~~~~~~~~~~~~~$\rx = 0$:~~~~~~Time Contraction (TC)
      \par
      ~~~~~~~~~~~~~~~~~~~~~~~~~~~$\rx'_0 = 0$:~~~~~~Length Dilatation (LD)
      \par
      ~~~~~~~~~~~~~~~~~~~~~~~~~~~$\rx_0 = 0$:~~~~~~Length Contraction (LC)
         \par The TC effect can be equivalently considered as TD in the reciprocal experiment.
           Langevin's arguments concerning RS and LC are based on consideration
      of the parametric equations for the $\rx_0 = 0$ case (Eqs. (2.17)-(2.19) above).
       However, to correctly describe events on the world lines of two spatially-separated
       clocks, Eq. (2.19) must be replaced
         by Eq. (2.46) with $\Delta t' = 0$ (no RS) to give (2.47) (no LC). In summary, only the
     projections $\rx = 0$ and $\rx' = 0$ of the generic LT, leading to time-like invariant interval relations,
    correctly describe the physical behaviour (the physically-independent TD effects in an experiment and its
     reciprocal) of actual synchronised clocks.
  
       \begin{table}
\begin{center}
\begin{tabular}{|c|c|c|c|c|} \hline
  Projection & $\rx' = 0$ &  $\rx = 0$ & $\rx'_0 = 0$ &  $\rx_0 = 0$ \\ \hline \hline
 `Effect' & $\rx_0 = \gamma \rx'_0$~~(TD) &   $\rx'_0 = \gamma \rx_0$~~(TC)  &  $\rx = \gamma \rx'$~~(LD) & 
   $\rx' = \gamma \rx$~~(LC) \\ \hline
 Parametric & $\rx = \gamma \beta \rx'_0$ &   $\rx' = -\gamma \beta \rx_0$  &  $\rx_0 = \gamma \beta \rx'$ & 
   $\rx'_0 = \gamma \beta \rx$ \\
 Equations & $\rx_0 = \gamma \rx'_0$ &   $\rx'_0 = \gamma \rx_0$ &  $\rx = \gamma \rx'$ & 
   $\rx' = \gamma \rx$ \\ \hline
  Hyperbola & $\rx_0^2-\rx^2 = (\rx'_0)^2$ & $(\rx'_0)^2-(\rx')^2 = (\rx_0)^2$ & $(\rx)^2-(\rx_0)^2 = (\rx')^2$ &
  $(\rx')^2-(\rx'_0)^2 = \rx^2$  \\ \hline
  Invariant & & & &  \\
  Interval & TL & TL & SL & SL  \\ \hline
\end{tabular}
\caption[]{{\it Different projections of the unidimensional coordinates of the
        generic LT (2.51) and (2.52) giving the space-time effects of:
       Time Dilatation (TD), Time Contraction (TC), Length Dilatation (LD) and 
      Length Contraction (LC)~\cite{JHFAJP1}. TL: Time-like invariant interval, SL: Space-like invariant interval.}}      
\end{center}
\end{table}
 \par Noting that, in the correct application of the LT to the description of
 synchronised clocks, only the TD and TC effects are genuinely physical, the RS and LC effects being
 spurious consequences of misapplication of the transformation equations, Langevin's discussion of the
 connection between simultaneity (and the lack of it) and causality is now critically examined.
  \par It is shown above that Langevin's assertion:
  \par  
  {\it `In the normal conception of time, simultaneity is absolute, and considered
      to be independent of the reference system.'}~~(italics in the original)
  \par holds true, not only in `mechanics', according to the GT, but also for `the transformation
    equations of electromagnetism', according to the LT (see Eq. (2.30)).
    Langevin argues that the commonsense notion of the frame-independence of simultaneity
   follows from the postulate of instantaneous action-at-a-distance. This is a logical {\it non sequitur}.
   There is no necessity that spatially separated events be causally connected. If two 
   synchronised clocks are spatially separated there is no reason whatever that any causal 
    connection should exist between the spatially-separated and simultaneous events corresponding
    to each clock registering the same epoch. The absolute nature of simultaneity (Eq. (2.30) above)
    is a simple prediction of the LT with no relation whatever to the speed of light signals
     or that of any other causal influence. Contrary then, to Langevin's assertions, there is no
     connection between the space-time geometrical predictions of the LT and the speed
    of light signals, in spite of the fact that space-like and time-like intervals may be 
    classified by considering hypothetical light signals passing between the spatial positions
    of the pair of events used to define the interval. The predictions of the LT would
     remain valid in a universe where no photons existed.
     \par Langevin next discusses a `perfect solid' in mechanics as a means to obtain 
     instantaneous action-at-a-distance, later pointing out that in the real world 
     such solids do not exist due to the finite speed of sound in any actual
      material. This detour adds nothing to the logical chain of his reasoning.
     \par Langevin contrasts, in the context of the GT (`mechanics') the invariant time
      interval between two events with the non-invariance of spatial intervals. 
      After noting that the non-simultaneous events corresponding to the dropping
    of objects through a hole in the floor of a moving cart are spatially-coincident for observers in the
     cart and spatially-separated for observers on the road, Langevin gives the usual
    definition of the spatial interval used to define the dimension of an object:
               \begin{quotation}
\textwidth 15cm   
  It is only in the case that the two events are simultaneous that the spatial
        separation has an absolute sense, and does not change with the choice of cordinate
        system. It follows immediately that the dimensions of an object, for example the
       length of a ruler, have an absolute sense and are the same for observers at rest, or
      in movement, relative to the object.
               \end{quotation}
\textwidth 16cm
   \par It is shown above that this statement holds true, not only in Galilean relativity,
        as asserted by Langevin, but also in special relativity for the {\it measured} length
       of an object, defined as the spatial separation of two simultaneous events, 
     in different inertial reference frames, as a consequence of the LT (2.26),(2.27)
      (Eq. (2.33) above).
   \par Langevin then remarks (as is obvious from inspection of Fig. 1, and previously exemplified
   by the objects dropped from the cart) that two non-simultaneous events may appear at the same
    position for one group of observers and at different ones for a second group in motion
    relative to the first.  Langevin's following statement that a time interval cannot be annuled
    by any choice of observation frame, although strictly true if the observer is
   assumed to have a time-like four vector, i.e. to have a non-vanishing mass, needs to be 
    qualified. As a consequence of the TD relation: $\Delta t = \gamma \Delta t'$ it can be asserted
    that:
   \par {\sl The finite time interval $\Delta t$ in the inertial frame S appears to an
       observer in a suitably chosen inertial frame S' to be  $\Delta t'$ where
     $\Delta t' < \epsilon$ for any finite value of $\epsilon$, no matter
      how small.}
   \par The physical content of this statment is encapsulated in the phrase: `Time stands still for a photon'.
   \par Langevin states that the asymmetry in the Galilean concepts of space and time is removed in
   special relativity, in that time intervals, like spatial ones, depend, in the latter, on the frame
   from which they are observed. This is a misleading statement in that in both Galilean and special
   relativity, spatial intervals corresponding to the dimension of a physical object, defined
    as $\Delta t = \Delta t' = 0$ projections, are frame invariant, whereas in special relativity,
    unlike in Galilean relativity, time intervals are frame dependent due the TD effect. Thus the 
     situation with respect to space and time intervals is more symmetric in Galilean relativity
     (they are both invariant) as compared to special relativity where only space intervals
      are invariant. The frame dependence of spatial intervals only occurs (in both  Galilean and
      special relativity) for the case of non-simultaneous events which does not correspond to the
       $\Delta t = 0$ projection that is correctly stated by Langevin to measure the length of
     a moving ruler.
     \par The following and important assertion of Langevin is that events which coincide both 
     temporally and spatially, do so when observed in all frames of reference. Although not explicity stated by
     Langevin, this holds for both inertial and accelerated reference frames. This fundamental
     postulate of relativistic physics is the basis for synchronisation of clocks at rest
     in different reference frames~\cite{JHFSTP1}. For example, in the generic LT (2.4)-(2.6) clocks in the
     frames S and S' have vanishing epochs (both record zero time) when the origins of
     the frames S and S' are in spatial coincidence. As discussed in Ref.~\cite{JHFSTL} such
     spatio-temporal coincidences define a `corresponding epoch' in the two frames. The
     same concept is the basis of the `system external' synchronisation procedure
     of Mansouri and Sexl~\cite{MS}.
     \par Langevin next discusses the physical implications of Einstein's second 
     postulate concerning the constancy of the speed of light, without however mentioning
     the evident contradiction with the aether-disturbance model of electromagnetic waves
     which he advocates. In this discussion, as is universal in
     text books and the literature,  Langevin does not discriminate between between the kinematical
     transformation of the velocity of a single object as specified in some inertial frame, 
     and the transformation of the relative
     velocity between two objects. Only the former is correctly given by the conventional
     relativistic parallel velocity addition relation (PVAR), the predictions of which
     are described by  Langevin:  
               \begin{quotation}
\textwidth 15cm
      A first group of observers see a light wave
          propagate in a certain direction with the speed 300,000 km/sec and see 
          another group of observers run after the wave with an arbitary speed and yet,
          for the second group, the light wave moves reative to it at the same
          speed of  300,000 km/sec.
               \end{quotation}
\textwidth 16cm
      \par This is indeed the prediction of the PVAR, but it does not correctly describe what would
      be seen by the second group of observers. Suppose that latter are at rest in the frame S'. The
      {\it relative velocity} of these observers and the light signal in the frame S is $c-v$. The 
     correctly transformed velocity of the light signal relative to the observers in the frame S' is then
      not $c$, as asserted by Langevin, but, as will be shown in the next section,
     $\gamma(c-v)$~\cite{JHFSTP3,JHFRECP}. This transformation
     relation will be crucial for the correct discussion of the exchange of light signals between the
     stationary and travelling twins in Langevin's thought experiment, to be discussed in the
     following section. Langevin correctly describes as `paradoxical' the situation in which the  
     second group of observers see the light wave move with the velocity $c$. Indeed it is, and
    the paradox is resolved by use of the correct relative velocity transformation formula.
   \par Langevin next introduces the category of pairs of space-like separated events and remarks, as
    discussed above in connection with Eq. (2.7) that the spatial interval between the events is 
     is greater than the transit time of a light signal between their spatial positions. Langevin
     then postulates that no causal influence can propagate faster than the speed of light,
     and correctly concludes, in this case, that space-like separated events are causally
     disconnected. The observation-frame-dependence of the time ordering of events, evident in 
     Fig. 2, then poses no problem for causality. If, on the other hand, causal influences
     could propagate faster than the speed of light, certain observers would be able to
     see the `effect' event occuring before the `cause' event, which is impossible. Langevin
     then concludes, that causality (defined as the necessity that all observers see the 
     effect as subsequent to the cause) requires that no signal can propagate faster than the
      speed of light  and that the infinite velocity assumed to be possible for forces in
      Newtonian mechanics is replaced in special relativity by a limiting signal velocity
      for all forces equal to that of the speed of light in free space. Indeed, Langevin's
      `transformation equations of mechanics' (the GT) are recovered from the
       `transformation equations of electromagnetism' (the LT)  in the limit $c \rightarrow \infty$.
      \par The logic of Langevin's argument is as follows: Given RS, superluminal signal 
      propagation results in causality violation for some observers. Therefore imposing
       causality forbids signal propagation at superluminal speed. This argument
      was first given by Einstein~\cite{Ein2}. The flaw in this reasoning
     is that when the LT is correctly interpreted, RS does not exist. There is therefore no 
     argument, based on causality, that no signal can propagate faster than the speed of
     light\footnote{The argument of Einstein for the impossibilty of superluminal
       signal propagation, assuming causality and the existence of RS was quoted by
        J.S.Bell in an essay discussing non-locality in quantum mechanics~\cite{JBell}.}.
     Assuming however the correctness of RS and concluding that no signal can
     propagate faster than light Langevin goes on to draw the conclusion that instantaneous
     action-at-a-distance, as in Newtonian gravity, is untenable whereas a finite
     propagation speed through a medium (the luminiferous aether) by local contact interations,
      as envisiged by Faraday, is favoured ---this in spite of the evident contradiction between
     such a hypothesis and Einstein's second postulate which lead him to abandon completely
    the aether hypothesis in the 1905 special relativity paper. 
     \par Langevin's assertion that the postulate that the force of gravity propagates at the speed
       of light is consistent with all astronomical observations is not true. As pointed
      out by Laplace~\cite{Laplace} such a retarded
       interaction results in an unbalanced torque and breakdown of Newton's third law in orbital
       motion under gravitational forces. Observations of the stability of the orbital
       motion of binary pulsars have given a lower limit on the the speed, $v_g$, of the gravitational
        force of $v_g > 2 \times 10^{10}c$~\cite{VanFlandern}. The impossiblity of stable
        circular Keplerian orbits of two oppositely charged objects under a mutual
          retarded electromagnetic interaction has been demonstrated in Ref.~\cite{JHFJMPA} as well
          as the all-orders special-relativistic correction to Kepler's third law in the case of an instantaneous
          interaction.
        It is also a prediction of QED that the intercharge force $\propto 1/r^2$, mediated by the exchange of
         space-like virtual photons, is instantaneous~\cite{JHFPS2}. This can be seen either
          by following Feynman~\cite{Feyn1}, and taking the Fourier transform of the
         momentum-space electron-electron scattering amplitude in the centre-of-mass (CM) frame
          or, more simply, from relativistic kinematics. In the overall CM frame the virtual
        photon, in the lowest order Feynman diagram, exchanges momentum, but no energy, between
         the scattered electrons. This implies according to the relativistic formula 
            connecting velocity, momentum and energy: $u = p c^2/E$ (see Eq.~(3.43) below) that the velocity of the
          virtual, photon is infinite, i.e. the corresponding interaction is instantaneous. This
          prediction is consistent with the results of a recently published experiments that
           measured the speed of magnetic or electric force fields close to their source~\cite{JAP1,JHFNRQED,Frascexpt}.
  \par Langevin next proceeds to describe the LC effect, which is a corollory of RS (see Fig~2 and
     Eqs. (2.17),(2.18)). As explained above, these spurious effects result from incorrect use of the generic
     LT (2.4)-(2.6) to describe a synchronised clock not at the position of the spatial origin of S'.
    \par Finally, in the passages translated above, Langevin, after introducing
     time-like separated events, describes correctly the TD
  effect of Eq. (2.10) that follows from the geometry of Fig. 1. It is demonstrated by
   Eq.(2.30) and shown in Fig. 3 that the TD effect is a universal one for spatially separated and sychronised
    clocks; i.e. there is no RS effect. However, when discussing the invariant time ordering of time-like 
     separated events, Langevin states, incorrectly, that the order could be inverted for an observer moving
     between the spatial locations of the events at a speed greater than that of light:
               \begin{quotation}
\textwidth 15cm
 It is seen immediately that such an
           inversion requires a speed greater than that of light for the second
            system of reference relative to the first.
               \end{quotation}
\textwidth 16cm
      \par  As mentioned above, the TD effect $\Delta t = \gamma \Delta t'$ shows that, although, for large enough $\gamma$, 
      (that is, with velocity very close to the speed of light) $\Delta t'$ may be made
      as small as desired for any finite value of $\Delta t$, setting $v>c$ gives an
      (unphysical) value of $\Delta t'$ that is imaginary. It does not change its sign. In order
      to invert a time interval for time-like separated events it is necessary to travel, not faster
       than the speed of light, but backwards in time. 
 
\SECTION{\bf{Langevin's discussion of relativistic differential aging}}
     This section begins with an English translation of Langevin's original presentation
    of the `twin paradox' thought experiment. In the remainder of the section, Langevin's description
    of the differential aging effect will be critically examined in the light of his misinterpretation
     of the predictions of the LT for pairs of space-like separated events, as explained in the previous
   section. In particular, the prediction of the differential aging effect given by the generic LT
    (2.3)-(2.5) and the conventional PVAR of special relativity will be contrasted with that
     of the LT (2.26),(2.27) that correctly describes a synchronised clock at an arbitary position
     in the frame S'. A translation of the passages of Ref.~\cite{Langevin} concerning differential
     aging and the twin paradox thought experiment follows:
               \begin{quotation}
\textwidth 15cm
       Consider an element of matter in arbitary movement, and the succession of events that
       constitute the history of that element  ---its world line.
    For two such events sufficiently close, uniformly moving observers who are 
      located successively at the two events may be considered as if at rest relative to the
        element, the change of velocity of the latter being negligible in the interval
     between the two events. For these observers, the time interval between the two events,
      an element of what we call the {\it proper time } of the element will be shorter
      than for any group of observers located in another uniformly moving reference system.
      If we take now two arbitary events in the history of our element of matter,
      their time interval as measured by observers in non-uniform motion who have constantly
       moved together with the element, will be, by integration of the previous result, shorter
       than for the uniformly moving reference system 
      In particular, this latter reference system can be such that the two events 
            considered could be situated at the same point, relative to which an element
        of matter has moved in a closed path, and returns to its original postion,
        due to its non-uniform movement. { \it For observers at rest relative 
        to that element of matter, the time which has elapsed between the beginning and end of the path,
         the proper time of the element of matter, will be shorter than for the observers in
         the uniformly moving reference system.} In other words, the element
        of matter will have aged less between the beginning and end of its path than if it
        had not been accelerated, if it had instead remained at rest in a uniformly moving
        reference system.
           It can be further said that it is sufficient to move around, to undergo
        accelerations,
       to age less quickly; we shall see in a moment how much one might hope to gain in this manner.
        Let's give some concrete examples. Imagine a laboratory at rest relative to the
      Earth, the movement of which may be considered a uniform translation, and in the laboratory two 
      perfectly identical samples of radium. Our knowledge of the spontaneous evolution of radioactive
      material allows us to state that, if the samples stay in the laboratory, they will lose their
      activity in the same way in the course of time and retain constantly equal activities.
       But now let us send one of the samples away from the laboratory at high speed and then
       bring it back. This requires that, at least at certain moments, the sample was accelerated.
        We can therefore state that, on return, its proper time between the beginning and the end of
       its excursion being less than the time between the same events seen by an observer at rest
       in the laboratory, it will have evolved less than the other sample and in consequence will
       be more active; it will have aged less having moved about in a more vigorous way. The calculation
       shows that to obtain a difference of $10^{-4}$ between the activities of the two samples, it
       would be necessary to maintain, during the separation, the speed of the travelling sample at
       about 4000 km/sec
       \par Before giving another specific example, let us look at this result in a different way.
       Suppose that two elements of matter meet for the first time, separate and come together again.
       We can state that observers at rest relative to one or the other during their separation
       will not measure the same time of separation, having not aged in the same way. It
        follows, from what is stated above, that the one having aged less is the one whose motion
        during the separation was furthest removed from uniform motion ---the one most strongly
        accelerated.
        This remark gives a way, for any of us, who is willing to devote two
      years of his life, of knowing how the Earth will be in 200 years time, to explore the
       future of the Earth, by making a jump forward, in the history of the latter, of two
       centuries ---corresponding in his own life to only two years--- but without any hope
       of return, without the possiblity to come and inform us of the result his journey,
       because any similar attempt at time-travelling can only throw him further and further
       into the future. 
   It is sufficient for this that our traveller agrees to shut himself up in
      a spaceship, sent away from the Earth, with a speed sufficiently close to that of light,
       although less (which is physically possible), arranging that an encounter occurs with,
       for example, a star, after one year in the life of the traveller, that sends the 
       spaceship back towards the Earth with the same speed. Returning to Earth, having 
       aged by two years, he will climb out of his vehicule and find our globe aged
       by at least 200 years if his speed had stayed within an interval of less than
       $5 \times 10^{-5}$ of the speed of light  Solidly established experimental facts of
       physics allow us to state that the situation will really be as just described. 
    It is amusing to realise how our explorer and the Earth see each other live,
     it they could, by light signals or wireless telegraphy, stay in contact during their
      separation, and to understand in that way how the asymmetry between the two measurements
      of the time of separation is possible.
  While they are separating from each other with a speed close to that of light,
     each one will seem to the other to run away from the electromagnetic or light signals
     which are sent towards him or her, so that it will take a very long time to receive the
      signals sent during a certain time interval. The calculation shows that each one of
      them will see the other live two hundred times more slowly than normal. During the year
     for which the motion lasts for him the explorer will receive from the Earth the news
       of only the first two days after his departure; during this year he will see just two
       days of Earth history. Besides, for the same reason, due to the Doppler shift,
      the radiation that he receives from the Earth during this time will have, for him,
       a wavelength two hundred times larger than on Earth. What seems to him visible light
       radiation, by which he could directly observe, by eye, the Earth, would be emitted
       by the latter as extreme ultra-violet radiation, similar perhaps to R\"{o}ntgen rays.
        If it is wished to maintain between them communication via hertzian signals, by
       wireless telegraph, the explorer having brought with him a receiving antenna
      of a certain size, the transmitter used on Earth during the two days following
       the departure of the explorer should be two hundred times shorter
       than his receiver.
      During the return journey the conditions will be inverted. Each one will
    see the other live a life markedly accelerated relative to his own, two hundred times
    times faster, and during the year which the return takes for him, the explorer
      will see the Earth accomplish the history of two centuries. It is thus understood
      how he will find it, on his return, aged by two hundred years. He will observe it by
       eye during this period by visible light waves, that are emitted from the Earth
      in the far infra-red region, with a wavelegth of about 100 microns, as recently
      discovered by Rubens and Wood in the emission spectrum of an Auer mantle.
     In order that he continues to receive from the Earth hertzian signals, these should,
      after the first two days, and during the two centuries which follow, use
      a transmitter  two hundred times longer than the traveller's receiver or
       forty thousand times longer that the one used during the first two days.
    To understand the asymmetry, it should be noted that the Earth will
     take two centuries to receive the signals sent out by the explorer during his motion
      away from the Earth, which takes, for him, one year. A terrestrial observer will
     see the explorer in his space vehicule live a life slowed down by a factor
      of two hundred; one year only of the explorer's life will be observed on
      Earth. During the two centuries during which the terrestrial observer will
      view the explorer receding, the former must, to receive the  hertzian signals
      sent by the latter, use a receiver two hundred times longer than the explorer's 
      transmitter. At the end of two centuries, arrives at the Earth the news of
      the encounter of the spaceship with the star, which marks the beginning of the 
       return voyage. The explorer will arrive at the Earth two days later, during
      which interval the terrestrial observer will see him live two hundred times
      faster than himself, thus seeing him accomplish the events of another year
      in order to find him aged, on return, by only two years. During these last two
      days, to receive news from the explorer, the terrestrial observer must use
      a receiver two  hundred times shorter than the explorer's
      transmitter.
     In this way, the asymmetry due to the fact that only the explorer has
      has undergone, in the middle of his journey, an acceleration which changes the
      direction of his velocity and sends him back to his starting point on Earth,
      results in the explorer seeing the Earth recede from him and approach him, during
      equal times, each of one year, while the  terrestrial observer, made aware of the
      acceleration only by the arrival of light waves, sees the explorer recede during
      two centuries and approach during two days, i.e. during a time
      forty thousand times shorter.
               \end{quotation}
\textwidth 16cm 
    At the beginning of the passages translated above, Langevin compares time intervals observed
       in the proper frame of an `element of matter' to those observed in frames that are in motion
       relative to the element. In order to give an operational meaning to the statements it is 
       necessary to assume that the different observers considered are equipped with clocks.
       Langevin's statement: 
      
               \begin{quotation}
\textwidth 15cm 
        {\it For observers at rest relative 
        to that element of matter, the time which has elapsed between the beginning and end of the path,
         the proper time of the element of matter, will be shorter than for the observers in
         the uniformly moving reference system.}~ 
       ( italics in the original)
               \end{quotation}
\textwidth 16cm 
          describes the situation in the primary experiment where the TD effect is given by Eqs.~(2.24) or (2.28).
          In this case the `element of matter' may be identified with one of the clocks
        C'$_1$ or C'$_2$ at rest in S'. However, as is clear from Eqs. (2.24) or (2.28), if the
        `element of matter' is instead identified with the clocks  C$_1$ or C$_2$ at rest in S, the time interval
         registered by these clocks, as seen by an observer at rest relative to them, is greater than, not less than, the
         corresponding time interval recorded by  C'$_1$ or C'$_2$ , which are in motion
         relative to the `element of matter' considered. Langevin's 
         statement is simply incorrect for observers at rest in S in the primary experiment. On the contrary,
         in the physically-independent reciprocal experiment  where the coordinates of the
         clocks  C$_1$ or C$_2$, not C'$_1$ or C'$_2$, appear in the LT, the time interval registered by
         the moving clocks C$_1$ or C$_2$ appears shorter for observers at rest in S'. 
          For some `groups of observers' then, the converse of
          Langevin's assertion is true, and the proper time interval in the rest frame of the `element of matter'
          may be greater than, not less than, the corresponding proper time interval registered in
          the proper frame of observers that are in motion relative to it.
        \par Langevin's subsequent assertion, offered as the explanation of the differential aging effect
        in the experiment of the twins, that `time passes more slowly in accelerated frames' is therefore
        not true in all cases. Indeed, in an instructive variant of the twins experiment~\cite{Halsbury,Marder}
       where no acceleration
        occurs, the travelling twin transmitting her clock setting to her uniformly moving older sister
         en route for the Earth, on arriving at the distant star, shows exactly the same differental
         aging effect as if she had herself rapidly declerated then accelerated back in the direction of the
         Earth at the end of her outward journey. In the version of the twins experiment proposed in
        Ref.~\cite{JHFSTP3} the travelling twin is also found to age less although both twins undergo
        identical impulsive acceleration programs. There is therefore no basis for Langevin's assertion
         that the differential aging effect is produced by the acceleration phases of the journey of
        the travelling twin. 
        \par Langevin introduces a sample of radioactive material (a `source') to constitute a clock, C. If the
        sample consists initially (at time $t = 0$) of $N_0$ undecayed nuclei, the exponential decay law
         $N(t) = N_0 \exp[-t/\tau_D]$ where $N(t)$ is the number of remaining undecayed nuclei at time $t$,
        and $\tau_D$ is the mean decay lifetime of the radioactive nucleus, gives a relation between
        $N(t)$ and and the epoch $t(\RC)$ of the equivalent clock:
          \beq
          t(\RC) = \tau_D\ln\left(\frac{N_0}{N(t)}\right)
          \eeq
          Langevin remarks that the activity of the sample, $A$,  (the number of disintegrations per
          unit time) also respects an exponential decay law:
          \beq
          A(t) \equiv -\frac{d N(t)}{dt} = \frac{N_0}{\tau_D} \exp\left[-\frac{t}{\tau_D}\right]
          \eeq
           so that 
             \beq
          t(\RC) = \tau_D\ln\left(\frac{A_0}{A(t)}\right)
          \eeq
          If the activity is measured by placing the sample in the same detector at different times
           the epoch $t(\RC)$ of C is given directly by substituting the measured activities in (3.3).
            Since $A_0^{meas} = \epsilon A_0$ and  $A(t)^{meas} = \epsilon A(t)$ the efficiency,
            $\epsilon$, of the detector cancels from the ratio of the measured activities:
           \beq
              \frac{A_0^{meas}}{A(t)^{meas}} =  \frac{\epsilon A_0}{\epsilon A(t)} =
                 \frac{ A_0}{ A(t)}
           \eeq
            Langevin discusses the differential aging, due to TD, of two such radioactive clocks, one at
            rest in the laboratory and the other undergoing a round trip like the travelling twin in the
             thought experiment that is subsequently proposed. The TD relation and Eq. (3.2) give, for a small
              difference, $\Delta A$, of the activities of a uniformly moving
             and stationary clocks
               \[ \frac{\Delta A}{A} = \frac{t}{\tau_D}\left(\frac{\gamma-1}{\gamma}\right)  \]
             If  $\Delta A /A = 10^{-4}$ at $t = \tau_D$ then $\gamma = 1.0001$, $\beta = 0.014c$ and $v = 4200$ km/s
             in agreement with Langevin's assertion. Note that the parameters of the acceleration 
           and deceleration program (as in the following analysis of the twins thought experiment) play no
            role in this calculation.
             \par Consideration of two such radioactive clocks, spatially-separated and undergoing
            identical motion, makes evident the absurd nature of the `relativity of simultaneity' effect
            of conventional special relativity theory. Langevin states, correctly, that 
       
               \begin{quotation}
\textwidth 15cm 
              `...if the samples stay in the laboratory they will lose their activity in the same way in
              the course of time and retain constantly equal activites.'
               \end{quotation}
\textwidth 16cm 
 It is obvious that if {\it both} samples move in an identical manner then, although their activities
      will both change relative to a sample at rest in the laboratory, they will do so in an identical manner. They will
     therefore, also in this case, `retain constantly equal activites.'
        If two similar radioactive sources are prepared simultaneously in the laboratory
            with the same value of $N_0$ they constitute a pair of naturally synchronised clocks. If they then 
           move in in a similar manner they will remain, because
           of the constancy of $\tau_D$,  synchronised at all later times, not only as 
            observed in their own proper frame, but also in the laboratory frame, since they both
            are subject, in this frame, to the same TD effect that has no dependence on the spatial position of
            the clocks. Suppose that they are both accelerated to speed $v$ in the laboratory, by
           an impulsive acceleration program occupying a negligibly short time in the laboratory.
           Calling the equivalent clocks $\RC_1$ and $\RC_2$, and a similar clock at rest in the
           laboratory frame $\RC_3$, the TD relation (2.10) gives:
          \ba
            t_3(\beta,t_1) & = & \gamma t_1 \\
            t_3(\beta,t_2) & = & \gamma t_2 
          \ea
             where $t_1 \equiv t(\RC_1)$ etc. Events which are simultaneous for the clocks
           $\RC_1$ and $\RC_2$:
          \beq
           t_1 = t_2 \equiv t
          \eeq
        are then also simultaneous as recorded by the clock
       $\RC_3$  since
           \beq
         t_3(\beta,t_1) = t_3(\beta,t) = t_3(\beta,t_2) =  \gamma t
          \eeq
           An experimental test of this prediction would be obtained my measuring the activity of the moving
             clocks by similar, and similarly placed, detectors in the laboratory frame ---such
            detectors are expected to measure equal activities at any instant--- as would be also
            measured by symmetrically placed detectors in the proper frame of  $\RC_1$ and $\RC_2$.
             For further discussion of this incompatiblity of `relativity
            of simultaneity' with the properties radioactive clocks see~\cite{JHFLLT,JHFSTP3,JHFSTL}.

              \par The travelling twin thought experiment will now be analysed, firstly in terms of
                   the postulates and concepts of standard special relativity theory (SSRT), as employed
                   in Langevin's account of the experiment, and secondly by use of the LT (2.39)-(2.40)
                   that correctly describes synchronised clocks at different spatial positions.
                   These analyses are based on different postulates or assertions selected from the following list:

               \begin{itemize}
                \item[(i)] The generic LT (2.4)-(2.6)
                \item[(ii)] The Parallel Velocity Addition Relation (PVAR):
                 \beq
                  w = \frac{u-v}{1-\frac{uv}{c^2}}
                  \eeq
                \item[(iii)] The number of light signals recorded by the travelling observer, T, or the
                     Earth-bound observer, E, at a corresponding epoch is the same whether calculated
                     in the proper frame, S, of E or the proper frames, S' or S'', of T.
 
                \item[(iv)] The LT (2.39)-(2.40), that correctly describes synchronised clocks at different
                          spatial locations.
                \item[(v)] The Relativistic Relative Velocity Transformation Relation (RRVTR),to be derived below:
                 \beq
                  u' = \gamma(u-v)
                 \eeq
                  \end{itemize} 
               Postulates among (i)-(iii) are used in the SSRT analyses, (iv) and (v) in what is claimed here
               to be the correct relativistic analysis.
              \par Langevin's traveller moves at a speed such that $1-\beta = 5 \times 10^{-5}$ corresponding
              to $\gamma = 100$, For clarity of the figures showing spatial dispositions of objects and
              light signals in different frames, it is assumed, in the following, that T moves at the more
               modest speed of $v = c/2$ corresponding to $\gamma = 2/\sqrt{3} = 1.155$. As in Langevin's
              discussion, only the phases of uniform motion of T are considered. On the outward journey, the
              proper frame, S', of T moves with speed $v$ along the line separating the Earth and the
              star Sirius (Si), which is at a fixed distance, $L$, from the Earth in the proper frame, S,
              of the Earth and Sirius. During the return journey, the proper frame S'' of T moves with
              speed $v$ in the direction from Sirius to the Earth. For reasons that will become
              apparent below, three further `mark' stars lying at fixed positions, in the frame S, on the
              line connecting Sirius to the Earth, are considered; S$_1$ is distant $2L$ from the
              Earth on the opposite side to Sirius, and  S$_2$ and  S$_3$ are at distances $2L$ and 
              $3L$ respectively, on the same side as Sirius. Both E and T are equipped with light signal
              transmitters that send short pulses, at a repetition frequency $\nu_0$, that are continously
              observed by the other twin during the voyage. In the frame S, successive light signals sent by
              E are separated by the distance $\delta_0 = L/10$. The first light signals of both
              transmitters are sent at the beginning of T's outward journey at epochs $t = 0$, $t'=0$ in S,S'. 
 \begin{figure}[htbp]
\begin{center}\hspace*{-0.5cm}\mbox{
\epsfysize18.0cm\epsffile{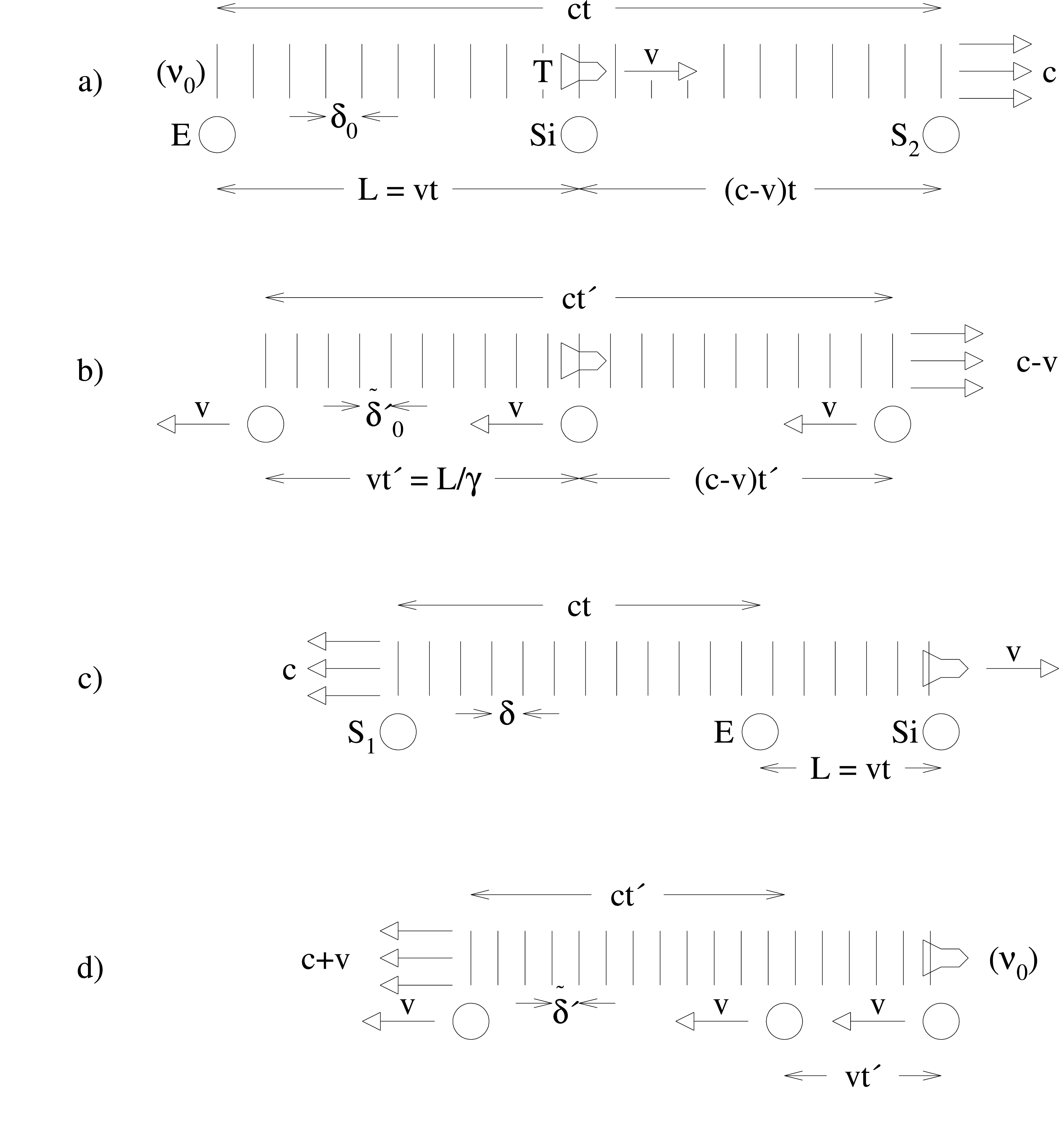}}
\caption{ {\em Spatial configurations of E,T,Si,S$_1$,S$_2$,and light signals at the end of T's
          outward journey. a) In S, the rest frame of E,Si,S$_1$ and S$_2$ showing light signals
         sent from the Earth at frequency $\nu_0$; b) in S', the rest frame of T, as calculated using
  postulates (i), (ii) for E,Si,S$_1$ and S$_2$, and (iii) for light signals; c) in S, showing all light signals
          sent from T with time-dilated frequency $\nu_0/\gamma$; d) in S', where T sends signals of
            frequency $\nu_0$, calculated as in b) above. $v=c/2$, $\gamma = 1.155$. See text for discussion.}}
\label{fig-fig5}
\end{center}
\end{figure} 
 \begin{figure}[htbp]
\begin{center}\hspace*{-0.5cm}\mbox{
\epsfysize18.0cm\epsffile{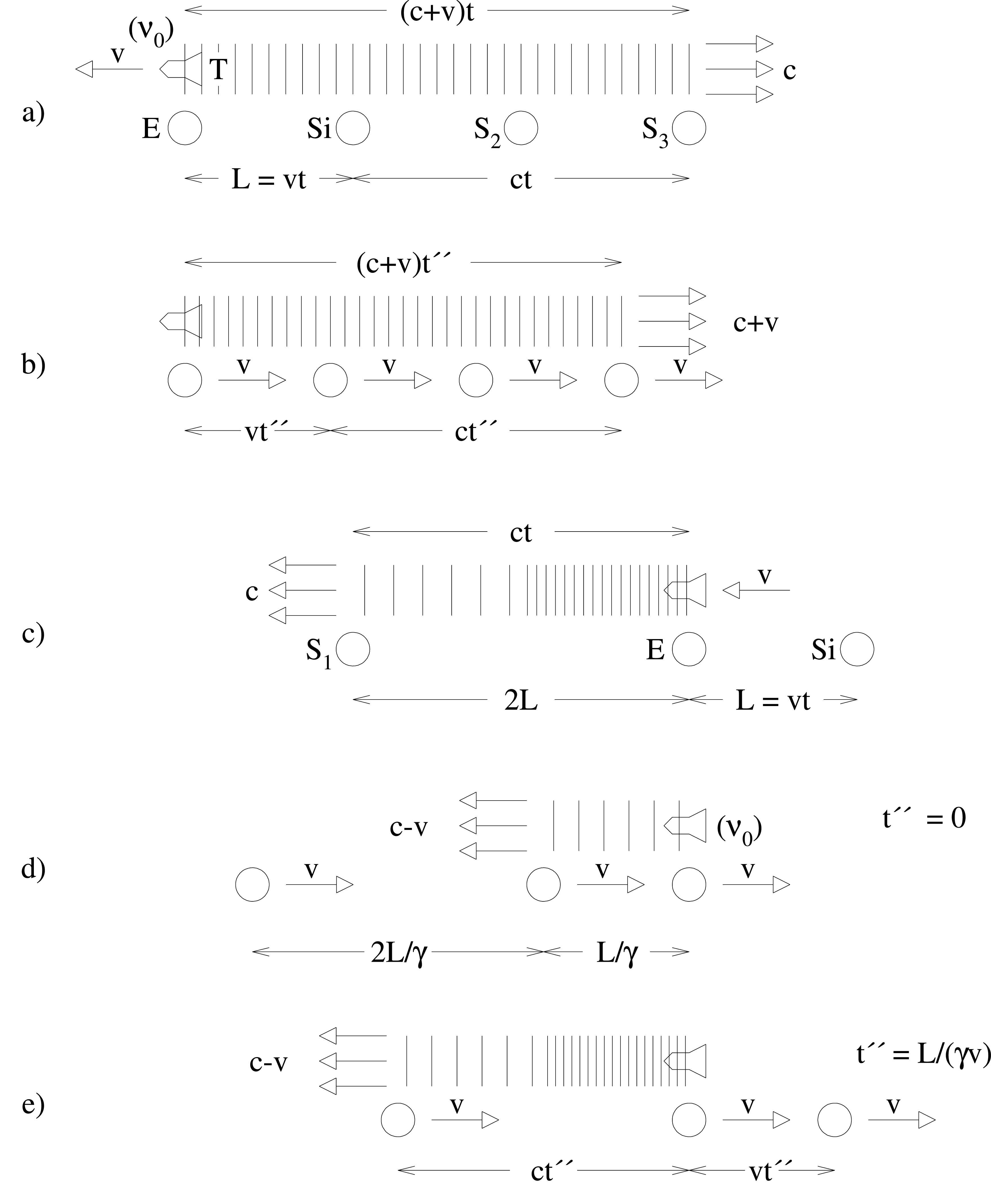}}
\caption{ {\em Spatial configurations of E,T,Si,S$_1$,S$_2$,S$_3$  and light signals during T's
          return journey. Clocks are reset to zero at the beginning of the journey.  a) In S, the rest frame of E,Si,S$_1$,S$_2$
    and S$_3$, at the end
            of the return journey, showing light signals
         sent from the Earth at frequency $\nu_0$. b) in S'',the rest frame of T, at the end
            of the return journey, as calculated using
  postulates (i), (ii) and (iii). c) in S, at the end
            of the return journey, showing the light signals, sent from T with time-dilated frequency $\nu_0/\gamma$ 
             that are received by E during the return journey; d) in S'', at the begining of the return journey, showing
             light signals sent by T at frequency $\nu_0$ during the outward journey, but not yet received by E, calculated
             as in b);
            e) in S'', at the end of the return journey, showing all light signals received by E during the
               return journey, calculated as in b). $v=c/2$, $\gamma = 1.155$. See text for discussion.}}
\label{fig-fig6}
\end{center}
\end{figure} 
              \par The configuration in S at the end of the outward journey at epoch $t =L/v$, of
               E, Si, S$_2$ and the light signals sent from E, is shown in Fig. 5a, that of  E, Si, S$_1$ 
               and the light signals sent from T, in Fig. 5c. It can be seen that the first signals
               from E and T are in spatio-temporal coincidence with the world lines of S$_2$ and  S$_1$ respectively.
               The observed frequencies of the E signals at T and the T signals at E are readily calculated
               from the geometries of Fig. 5a and 5c respectively. T receives the number $[(c-v)/c]\nu_0t$ 
                of signals during the time interval $t'$ corresponding to an observed frequency:
               \beq
          \nu_T^O(S) = \frac{(c-v)}{c}\frac{\nu_0t}{t'} = (1-\beta)\gamma \nu_0 = \sqrt{\frac{1-\beta}{1+\beta}}\nu_0
               \eeq
      where the TD relation $t =\gamma t'$ has been used. In the symbol $\nu_a^b(F)$ on the left side
         of Eq. (3.11), $a$ specifies the observer ($T$ or $E$), $b$ the phase of the journey ($O$, outward; 
          $R$, return) and $F$ the frame ($S$, $S'$ or $S''$) in which the calculation of the frequency
          is performed. A similar calculation for E gives:
               \beq
          \nu_E^O(S) = \frac{c}{(c+v)}\left(\frac{\nu_0t}{\gamma}\right)\frac{1}{t}= 
            \frac{\nu_0}{(1+\beta)\gamma} = \sqrt{\frac{1-\beta}{1+\beta}}\nu_0
               \eeq
              Due to TD, the frequency of the signals from T is reduced by the factor $1/\gamma$ in the frame S. 
              Therefore  $[c/(c+v)](\nu_0/\gamma)t$  signals are recorded by E during the time interval $t$,
              leading to the result of Eq. (3.12).
             For Langevin's travelling twin with $\gamma = 100$ the reduction factor $\sqrt{(1-\beta)/(1+\beta)}$
              takes, to a very good approximation, the value $1/200$,  so that as correctly stated by Langevin:
               \begin{quotation}
\textwidth 15cm \
   .~.~.~each of them will see the other live two hundred times more slowly \newline than usual.
               \end{quotation}
\textwidth 16cm
            \par The observed frequencies during the outward journey are now calculated in the proper frame S'
            of T according to SSRT, using postulate (i) to calculate the positions
               of the astronomical objects in S', (ii) to calculate their velocities
            in S' and (iii) to deduce the positions of the light signals in S' from those shown in S in
              Figs. 5a and 5c. Transforming the coordinates of E and Si into the frame S' at $\rt = 0$ when
             $\rx_E = 0$, and $\rx_{Si} = L$, using the generic LT (2.4)-(2.6) gives:
              \ba 
               \rx'_E(\rt'_E(\rt)) & = & \rx'_E(0) =  0,~~~~\rt'_E(\rt) = \rt'_E(0) = 0 \\
      \rx'_{Si}(\rt'_{Si}(\rt)) & = &    \rx'_{Si}(\rt'_{Si}(0)) = \gamma L,
       ~~~~\rt'_{Si}(0) = \frac{\gamma v L}{c^2}
               \ea
           According to Eq. (3.9), E moves with velocity $v$ in the negative $\rx$-direction in S' ( $w= -v$ on
       setting $u = 0$). It follows that at $\rt' = 0$ the $\rx'$ coordinate of Si is:
         \beq
          \rx'_{Si}(0) = \gamma L - \left(\frac{\gamma v L}{c^2}\right)v  = \frac{L}{\gamma} 
         \eeq
        Thus the distance of Si from E at  $\rt' = 0$ is:
          \beq
          \rx'_{Si}(0) - \rx'_{E}(0)  =  \frac{L}{\gamma}  
         \eeq
          which is the familiar `length contraction' effect. Note that it is a corollary of the 
          `relativity of simultaneity' effect in the frame S' shown in (3.13) and (3.14).
          A similar calculation shows that 
              \beq
          \rx'_{S_2}(0) - \rx'_{E}(0)  =  \frac{2L}{\gamma}  
         \eeq
           Since E, Si and S$_2$ are predicted by Eq. (3.9) to move to the left with speed $v$ in S',
            the spatial configuration in this frame at the end of the outward journey is that shown
           in Figs. 5b and 5d. According to postulate (iii) the number of light signals having crossed any object
           must be the same at corresponding epochs in the frames S and S'.
      It follows that the light signals in spatio-temporal 
            coincidence with E, Si, S$_1$ and  S$_2$ in S in Figs. 5a and 5c must also be so in S'
          at the corresponding epoch, as shown in  Figs. 5b and 5d, when $\rt = L/v$, $\rt' = L/\gamma v$.
          It can be seen that the spatial configurations in S' of  Figs. 5b and 5d are obtained from those
          in S in  5a and 5c by simply scaling all spatial dimensions by the `length contraction' factor
          $1/\gamma$. Denoting by $\tilde{\nu}$ frequencies calculated in S' according to the
           above prescription (i.e. SSRT with the velocities of the astronomical objects calculated
           according to (3.9) and the positions of the light signals according to postulate (iii))
           the geometry of Fig. 5b gives the same result for the frequency observed by T as the S-frame
         calculation of (3.12) since the same number of signals $[(c-v)/c]\nu_0\rt$ is recorded during
         the same time interval $\rt'$, hence:
                       \beq
         \tilde{\nu}_T^O(S') =   \nu_T^O(S) 
               \eeq
           In Fig. 5d, the number of signals received by E during the time interval $\rt'$ is $[c/(c+v)]\nu_0\rt'$
            corresponding to a frequency in the frame S' of $\nu'= \nu_0/(1+\beta)$. Since the TD relation
           gives $\nu' = \gamma \nu$, the frequency observed by E in the frame S is:
            \beq
         \tilde{\nu}_E^O(S') = \frac{\nu'}{\gamma} = \frac{\nu_0}{\gamma(1+\beta)} = \sqrt{\frac{1-\beta}{1+\beta}}\nu_0              \eeq
        so that comparing with Eq. (3.12) gives
         \beq
         \tilde{\nu}_E^O(S') =   \nu_E^O(S) 
               \eeq
         Consistent results for both observed frequencies are therefore given by calculations performed in 
        the frames S and S'
            \par The calculation in the frame S of the frequency observed by T during the return journey,
           based on the geometry of Fig. 6a, and in the frame S' based on the geometry of Fig. 6b, proceeds in
            the same manner as for the outward journey with the replacement $v \rightarrow -v$ in the equations
           yielding the results:
              \beq
         \tilde{\nu}_T^R(S') = \nu_T^R(S) = \sqrt{\frac{1+\beta}{1-\beta}}\nu_0
                \eeq
          \par As pointed out by Langevin, the calculation of the frequency of the signals received by E during
         the return journey is complicated by the fact that at the start of the return
         journey~\footnote{For simplicity, the clocks registering the epochs $t$ and $t''$ are set to 
            zero at the beginning of the return journey.}( $t = t'' = 0$) some light signals sent by T during the
           outward journey have not yet arrived at E (see Fig. 6d). Referring to Fig. 5c, since the spatial
          separation in S of the signals from T is:
              \beq  
                   \delta = \frac{c}{\nu} = \frac{c}{\nu_0} \sqrt{\frac{1+\beta}{1-\beta}}
               \eeq
              the number of signal intervals recorded by E during the outward journey is:
                \beq  
       N^{int}_E(\rm{obs})  = \frac{c t}{\delta} =  \sqrt{\frac{1+\beta}{1-\beta}}N^{int}_0
               \eeq
         where $N^{int}_0 = \nu_0 t$ is the total number of signal intervals sent by E during either the outward or the
         return journey. The number of signal intervals sent by T, but not yet observed by E, at the beginning of
         the return journey is
               \beq  
       N^{int}_E(\rm{non- obs})  = \frac{v t}{\delta} =  \beta \sqrt{\frac{1+\beta}{1-\beta}}N^{int}_0
               \eeq
         For the choice of parameters of Fig. 5: $\beta = 1/2$, $N_0 = 20$,
         $N^{int}_E(\rm{obs}) = N^{int}_0/\sqrt{3} = 11.55N_0$
            and $N^{int}_E(\rm{non-obs}) = N^{int}_0/2\sqrt{3} = 5.77N_0$. Since the number of received signals is one
           greater that the number of complete signal intervals received, E receives 12 signals during
        the outward journey and, at the end of it, 6 signals sent by T have not yet been received by E.
      The non-observed signals
        during the outward journey are recorded by E during the return journey with freqency  $\nu_E^O(S)$.
        The frequency of the signals sent by T and received by E during the return journey is calcuated
        as in (3.12) with the replacement $v \rightarrow -v$ to give:
     \beq 
      \nu_E^R(S) = \sqrt{\frac{1+\beta}{1-\beta}}\nu_0
           \eeq
        During the return journey, signals of frequency $\nu_E^O(S)$ are observed during the time interval
        $\beta t$ and signals of frequency $\nu_E^R(S)$ during the time interval $(1-\beta)t$. The ratio of the numbers
        of signals observed is therefore 
              \beq
    \frac{N_E[\nu_E^O(S)]}{N_E[\nu_E^R(S)]} = \frac{\beta\nu_E^O(S)}{(1-\beta)\nu_E^R(S)}
          = \frac{\beta}{(1-\beta)} \sqrt{\frac{1-\beta}{1+\beta}}\sqrt{\frac{1-\beta}{1+\beta}}
           =  \frac{\beta}{(1+\beta)}
        \eeq
       For $\beta = 1/2$ (see Fig. 6c) this ratio is $1/3 = 6/18$ whereas for Langevin's choice
       $1-\beta = 5 \times 10^{-5}$, $\beta \simeq 1$ it is, to a very good aproximation, equal to $1/2$.
        Thus during 99.995yr E receives signals at frequency $\nu_E^O(S)$ while in 0.005yr (1.83 days) he
          will receive twice as many signals at frequency $\nu_E^R(S)$, in agreement with Langevin's
          description.
          \par The above calculation shows agreement with the frequencies of the received signals, and 
          the time intervals over which they are observed, as given by Langevin. However, Langevin also
         asserts that the wavelength of the radiation received by T from the Earth during the outward journey
         is 200 times longer than its wavelength in the Earth's proper frame. If therefore
         $\tilde{\delta}'$ is the spatial separation in S' of the signals from the Earth on the outward
         journey, and it may be associated to a `wavelength', as suggested by the 
          the relation, valid in S, $c = \nu_0 \delta_0$, then Langevin claims that
          \beq
           \frac{\tilde{\delta}'}{\delta_O} = \sqrt{\frac{1+\beta}{1-\beta}} = 200~~~({\rm Langevin}) 
           \eeq
           whereas it is clear from inspection of Figs. 5a and 5b that $\tilde{\delta}'/\delta_O = 1/\gamma = 1/100$
           for Langevin's choice of $\beta$. The reason for this difference is that Langevin is assuming that the
           speed of the light signal in S' is $c$, as predicted by Eq. (3.9) whereas it is clear from inspection
           of  Figs. 5 and 6 that, under the assumptions of the calculation on which they are based, the speed 
          of the light signals from the Earth in T's proper frame is $(c-v)$ on the outward journey and
          $(c+v)$ on the return one. If the speed of light is the same in S and S', as assumed by  Langevin,
         the relation $v =\lambda \nu$ implies that the ratio of  wavelengths is, as asserted by  Langevin,
       the recipocal of the ratio of frequencies. However, on the outward journey $\tilde{\delta}'$ is actually given
         by the relation
        \beq
         \tilde{\delta}'_O = \frac{c-v}{\nu^O_T(S)} = c\frac{c-v}{\nu_0}\sqrt{\frac{1+\beta}{1-\beta}}
                           = \frac{\delta}{\gamma}
         \eeq
         and on the return journey by
        \beq
         \tilde{\delta}'_R = \frac{c+v}{\nu^R_T(S)} = c\frac{c+v}{\nu_0}\sqrt{\frac{1-\beta}{1+\beta}}
                           = \frac{\delta}{\gamma}
         \eeq
      Langevin assumes that the PVAR Eq. (3.9) which is used to calculate the velocity the massive
      astronomical objects E, Si, S$_1$,
    S$_2$ and S$_3$ in S' and S'' is also valid for the transformation of the velocity of the light signals
          from S into S' or S. Instead, the calculation, the results of which are shown in Fig. 5 and 6, fixes
       the positions of the light signals in S' and S'' from the known ones in S by using postulate (iii), that
      the same numbers of signals have been encountered by all objects at corresponding epochs in S and S' or S and  S''.  
       For example, since the simultaneous spatio-temporal coincidences of light signals with E, Si and S$_2$
       occurs at the epoch of the end of the journey in the frame S, the same coincidences
       must occur, as shown in Fig. 5b, at the corresponding epoch in the frame S'. 
  \begin{figure}[htbp]
\begin{center}\hspace*{-0.5cm}\mbox{
\epsfysize12.0cm\epsffile{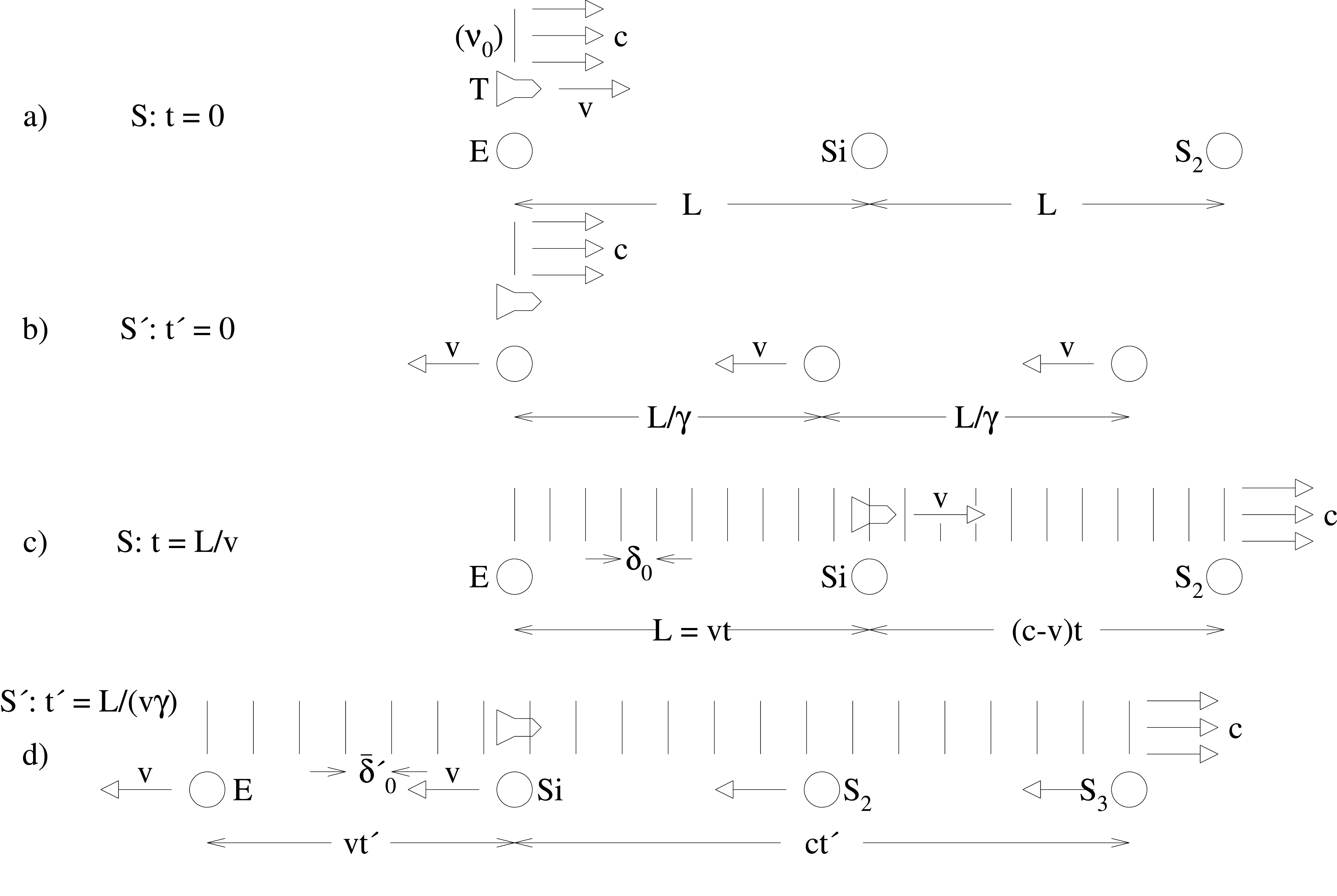}}
\caption{ {\em Spatial configurations of E,T,Si,S$_2$,S$_3$ and light signals sent from the Earth at frequency $\nu_0$
          during T's outward journey. a) In S, the rest frame of E,Si,S$_2$ and S$_3$ at the beginning
           of the outward journey; b) in S', the rest frame of T, at the beginning of the outward journey; as calculated using
  postulates (i), and (ii) for E,Si,S$_2$, S$_3$, and the light signals; c) in S,at the end of the journey (same as Fig.~5a);
        d) in S', at the end of the journey, calculated as in b) above. $v=c/2$, $\gamma = 1.155$. See text for discussion.}}
\label{fig-fig7}
\end{center}
\end{figure} 
        \par If it is assumed, with Langevin, that the PVAR gives correctly the velocity transformation 
         of both the massive astronomical objects and the light signals, it is found to be impossible to
        respect the condition (iii) for the light signals. This is illustrated in Fig. 7, 
        where the results of the calculation using the postulates (i) and (ii) both for massive objects
       and light signals is shown. The case of a light signal sent by E and received by T during the 
       outward journey is considered. Fig. 7a shows the configuration in S at the epoch, $t= 0$, when
       the first light signal is sent. The corresponding configuration in S' as  calculated from postulates
       (i) and (ii) is shown in Fig. 7b. At the end of the journey, the configuration in S, shown in Fig. 7c, 
        is identical to that of Fig. 5a, however that in S' at the corresponding epoch, shown in Fig. 7d,
         is markedly different from that of Fig.~5b. The first light signal from E is in spatio-temporal
           coincidence with S$_2$ in the frame S but in spatio-temporal coincidence with S$_3$ in the
          frame S'. Alternatively imposing the condition (iii) between the frames S and S' for the first
         light signal as is done in Fig. 5b, requires that the speed of the signal in S' is $c/2$, not $c$,
        in contradiction with the postulate (ii) applied to light signals. Futhermore, comparison of Figs. 7c
        and 7d shows that the condition (iv), that the number of signals recorded by T during the
        journey is the same, whether calculated in S or S' is not respected. With the choice of
        parameters in Fig. 7, ten signals are recorded in the frame S and fourteen in the frame S'.
        \par Calculating, in the frame S', the frequency $\bar{\nu}_T^O(S')$ of the signals from Earth 
        received from T according to the geometry of Fig. 7d, gives the result:
         \beq
    \bar{\nu}_T^O(S') = \frac{v}{v+c}\frac{\nu_0 t}{t'} = \frac{\gamma \nu_0}{1+\beta} = 
         \frac{\nu_0}{(1+\beta)\sqrt{1-\beta^2}} 
      \eeq
     which is not equal to $\nu_T^O(S)$. Comparison of the geometries of Fig. 7c and 7d shows that, if
      $\bar{\delta}'$ is the spatial interval between light signals in Fig. 7d,
      \ba
      \frac{\bar{\delta}'}{\delta_0} & = & \frac{{\rm Separation~of~E~and~}{\rm S}_3{\rm~in~Fig.~7d}}
       {{\rm Separation~of~E~and~}{\rm S}_2{\rm~in~Fig.~7c}} \nonumber \\
        & = & \frac{(v+c)t'}{ct} = \frac{1+\beta}{\gamma} = (1+\beta)\sqrt{1-\beta^2}   
      \ea 
       For the parameter choice of Fig.7, $\beta = 1/2$, it is found that       
 $\bar{\delta}'/\delta_0 = 3\sqrt{3}/4 = 1.30$, while for Langevin's choice of $\beta$, (3.31) gives 
       $\bar{\delta}'/\delta_0 = 1/50$, to be compared with Langevin's assertion in (3.27)
       $\tilde{\delta}'/\delta_0 = 200$. Since $c = \delta_0  \nu_0$, the frequency of (3.30) and
       the spatial separation $\bar{\delta}'$ given by (3.31) satisfy the condition $c = \bar{\nu}_T^O(S')\bar{\delta}'$. 
        Langevin assumed instead, in (3.27), that the relation $c = \nu_T^O(S)\tilde{\delta}'$ holds. 
       However, as shown in Fig. 7, application of the PVAR consistently to both massive 
       objects and light signals gives a value of the frequency $\bar{\nu}_T^O(S')$ 
       different from $\nu_T^O(S)$ as
       calculated in the frame S ---that is, inconsistent results.
       \par To summarise the comparison of the results shown in Figs. 5, 6  and 7: the calculation
        based on the postulate (i), postulate (ii) for massive objects only, and postulate (iii)
       to determine the position of light signals in S' and S'' from those calculated in S
       by consideration of spatio-temporal coicidences with the mark stars S$_1$, S$_2$ and S$_3$
      leads to consistent results for the observed freqencies, as calculated in different frames, that
      are also consistent with the well-known relativistic Doppler shift formula as derived by
      Einstein, to be discussed further below.
       However, inspection of Figs. 5 and 6 shows that the velocity of light
     signals from the Earth observed in the rest frame of T is $c-v$ during the outward journey and $c+v$ during the 
    return journey and that light signals from T received by E have speeds, in the rest frame
    of T, of $c+v$ during the outward journey and $c-v$ during the return journey. These
    transformed velocites are those given by the classical velocity addition relation
    $w = u-v$ and are in evident disagreement with the predictions of the relativistic PVAR (3.9).
     On the other hand, if the postulates (i) and (ii) are consistently applied to both massive
     objects and light signals, then, as shown in Fig. 7, inconsistent predictions are given by
     calculations performed in different frames. The frequency  $\bar{\nu}_T^O(S')$ calculated in S'
     differs from that $\nu_T^O(S)$ calculated in S. In this case also, the assertion 
      (iii) is negated. At the end of the outward journey the first light signal 
     sent from the Earth is in spatio-temporal coincidence with S$_2$ in the frame S (Fig. 7c)
     and with  S$_3$ in the frame S' (Fig. 7d). Inspection of Fig. 7 shows that different
    numbers of light signals are recorded by T in the frames S and S' at the end of
    the outward journey.
       \par These contradictory results show that incompatible postulates
    are used in the calculations so that, by {\it reductio ad absurdum}, one, or more,
    of these postulates must be
    false. If (i),(ii) for massive objects and (iii) are assumed (ii) is falsified for
     light signals. If (i) and (ii) are consistently applied to all objects, independently
     of their mass, postulate (iii) is falsified. Since postulate (iii) is equivalent to assuming
     invariance of the phase of a wave train ---as used by Einstein in his original
    derivation of retativistic Doppler-shift formulas--- it is clear that
     there is an essential incompatiblity between this postulate and the predictions
     of the generic LT of Eqs.(2.4)-(2.6) (postulate (i)) from which the PVAR (postulate (ii))
     has been be derived. A clue to a possible resolution of the problem is given by noting 
     that the length contraction effects of Eqs. (3.16) and (3.17), shown in the previous
     section to be spurious, play an essential role in obtaining consistent and correct
     (according to the kinematical derivation to be presented below)
      frequency predictions in different inertial frames.
 \begin{figure}[htbp]
\begin{center}\hspace*{-0.5cm}\mbox{
\epsfysize20.0cm\epsffile{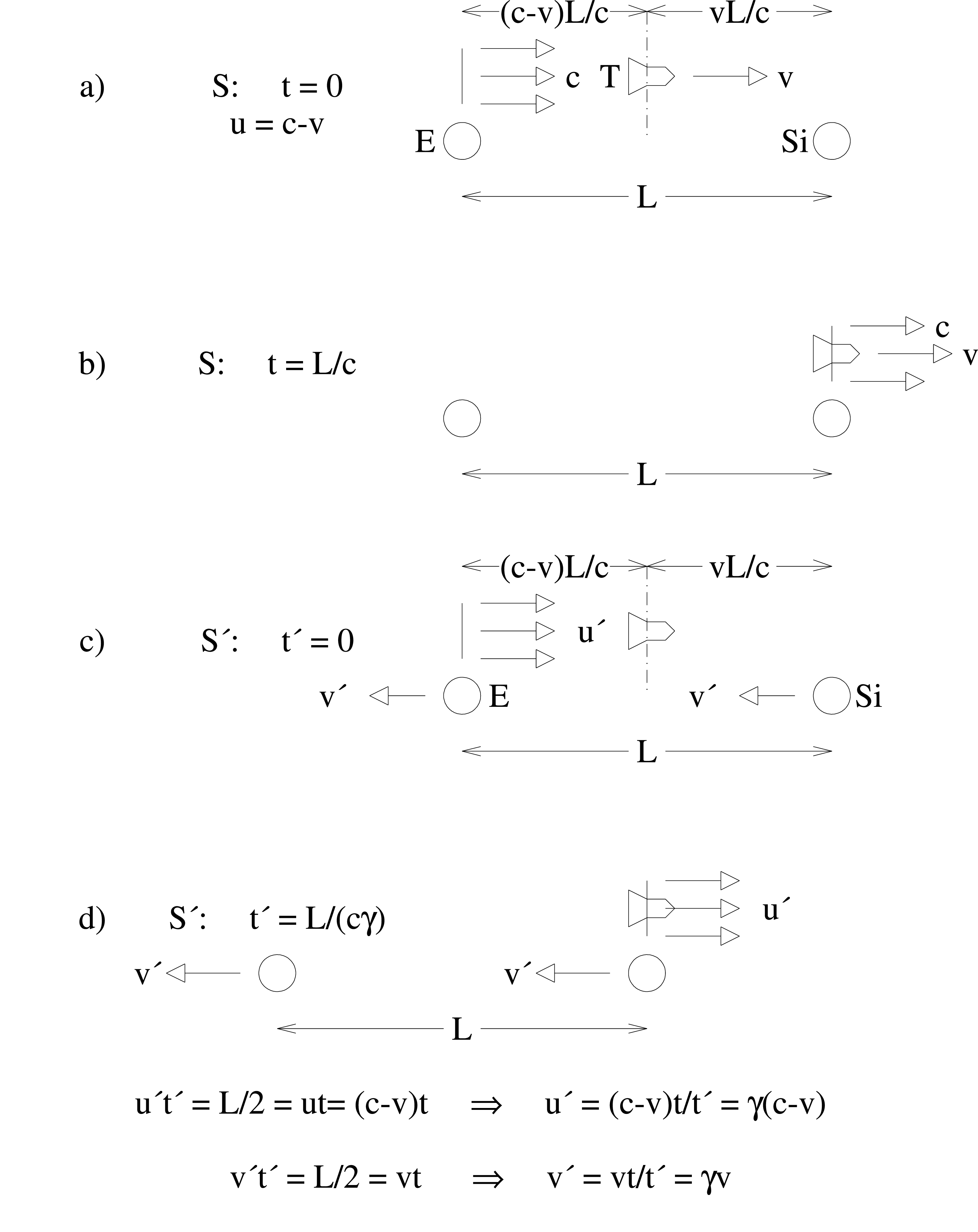}}
\caption{ {\em Spatial configurations of E,T,Si and a light signal during T's
          outward journey used to derive the RRVTR Eq.~(3.33).
      a) In S, the rest frame of E and Si, at $t = t' = 0$. The light signal
        is sent from the Earth when T is distant $(c-v)L/c$ from E. b) in S, at $t = L/c$ when the
           light signal,T and Si are in spatial coincidence. c) in S', the rest frame of T. at $t' = 0$.
         d) in S', at $t' = L/(\gamma c)$. As shown, the geometries of the configurations in c) and d) yield
           the RRVTR: $u' = \gamma(c-v)$,~$ v'= \gamma v$. $v=c/2$, $\gamma = 1.155$. See text for discussion.}}
\label{fig-fig8}
\end{center}
\end{figure}   

     \par Before showing the results of the calculations based on the postulates (iv) and (v),
      the Relativistic Relative Velocity Transformation Relation (RRVTR) of Eq. (3.10) is derived.
      For this, it is convenient to consider a configuration during the outward journey at
      epoch $t = 0$ shown in Fig. 8a, when T is
     distant $(c-v)L/c$ from E and a light signal is sent by E in the direction of
      Sirius\footnote{To simplify the equations, clocks in S and S' are reset to zero
       at the instant of the configuration shown in in Fig. 8a}. At
      epoch $t = L/c$ as shown in  Fig. 8b, there is a triple spatio-temporal coincidence between the
       world lines of
      T, Si and that of the light signal. The velocity relative to T of the signal in the frame S is
      $u = c-v$. The corresponding configurations in the frame S' at epochs $t' = 0$,
       $t'= L/(\gamma c)$ are shown in Figs. 8c and 8d respectively. Due to the invariance 
       of the separation of E and Si (see Eq. (2.33) above)
       the velocity $u'$ of the light signal relative to T in S' (the rest frame of the latter)
        is given by the geometry of Figs. 8c and 8d as
       \beq
        u' = \frac{(c-v)L}{ct'} = \frac{(c-v)t}{t'} = \gamma(c-v)
        \eeq
       while $v'$, the velocity of E or Si relative to T in S', is found to be
    \beq
        v' = \frac{vL}{ct'} = \frac{vt}{t'} = \gamma v
        \eeq
        Eq. (3.32) is the RRVTR of Eq. (3.10) for the special case $u = c$, while (3.33) is
        the special case of the relation for $u = 0$, noting the opposite directions of $v$ and $v'$ in
        Figs. 8a and 8c.
        \par The results of the calculations of configurations using postulates (iv) and (v)
         are shown in Fig. 9 for the frames S and S'  at the end of the outward journey
         and in  Fig. 10 for the frames S and S'' at the end of the return journey. It can be seen
        that the geometical configurations in the frames S' and S'' are the same as in Figs. 5 and 6
      except that there is no length contraction effect, while the velocities of the light
       signals in these frames are given by the RRVTR of Eq. (3.32) rather than by the
        classical relation $u' = u = c-v$.  
        \par Under these circumstances the calculations, in the frames S'and S'', of the
         frequencies of the signals recorded by T or E are algebraically identical to that
         of $\tilde{\nu}_a^b(F)$ discussed above. For example, the number of signals
        recorded by E in S' is $[c/(c+v)]\nu_0 t'$ in both Fig. 5d and Fig. 9d, so 
        that the same observed frequency is found. In consequence, if $\nu_a^b(S')$  and
         $\nu_a^b(S'')$ denote frequencies calculated on the basis of the geometry
        of Figs. 9 and 10 it follows that:
          \beq
   \nu_E^O(S')= \tilde{\nu}_E^O(S') =  \nu_E^O(S) =  \nu_T^O(S')= \tilde{\nu}_T^O(S') =  \nu_T^O(S)
       = \sqrt{\frac{1-\beta}{1+\beta}}\nu_0 
          \eeq
         while the corresponding frequencies for the return journey are given by thr replacements
         $O \rightarrow R$, $S' \rightarrow S''$  and $\beta \rightarrow -\beta$ in Eq.(3.34).
         \par The crucial difference between the configurations of Figs. 5 and 6 and those of
        Figs. 9 and 10  is the different physical explanation they give of the TD effect.
        Consider the motion of T between E and Si on the outward journey. In Fig. 5b, TD 
        results from `length contraction' in the frame S':
            \beq
       t' = \frac{L'}{v'} = \left(\frac{L}{\gamma}\right)\frac{1}{v} = \frac{t}{\gamma}
            \eeq
         whereas in Fig. 9b, it results from the greater relative velocity of Si and T in the
        frame S' than in the frame S:
          \beq
       t' = \frac{L'}{v'} = \frac{L}{(\gamma v)} = \frac{t}{\gamma}
            \eeq
          However, unlike the calculation based on the postulates (i), (ii) for massive objects
         and (iii) for photons, the results of which are shown in Figs. 5 and 6, that based
      on postulates (iv) and (v), shown in Figs. 9 and 10, although giving identical
       predictions for the observed frequencies, is entirely free of internal contradictions.
        It is therefore asserted here that these are the correct relativistic calculations.
       Only TD is then a {\it bona fide} physical effect. Indeed, it is the only one that
       is confirmed by experiment~\cite{JHFLLT,JHFSEXPS}. 
 \begin{figure}[htbp]
\begin{center}\hspace*{-0.5cm}\mbox{
\epsfysize18.0cm\epsffile{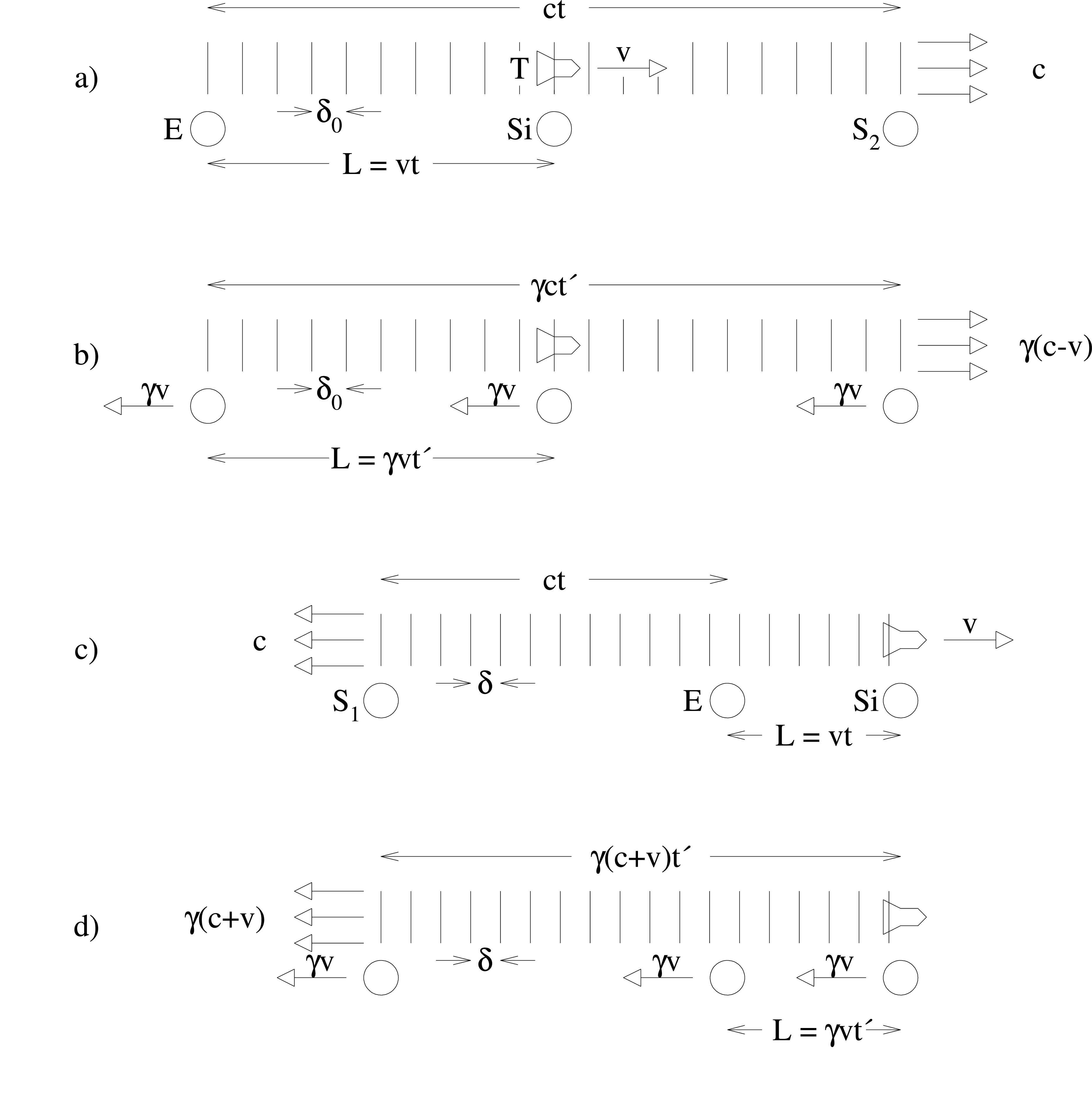}}
\caption{ {\em Spatial configurations of E,T,Si,S$_1$,S$_2$ and light signals at the end of T's
          outward journey. a) In S, the rest frame of E,Si,S$_1$ and S$_2$ showing light signals
         sent from the Earth at frequency $\nu_0$ (the same as Fig.~5a and Fig.~7c). b) in S', the rest frame of T, as calculated using
  postulates (iv) and (v). c) in S, showing light signals
          sent from T with time-dilated frequency $\nu_0/\gamma$ (the same as Fig.5c). d) in S', where T sends signals of
            frequency $\nu_0$, calculated as in b) above. $v=c/2$, $\gamma = 1.155$. See text for discussion.}}
\label{fig-fig9}
\end{center}
\end{figure}
  \begin{figure}[htbp]
\begin{center}\hspace*{-0.5cm}\mbox{
\epsfysize18.0cm\epsffile{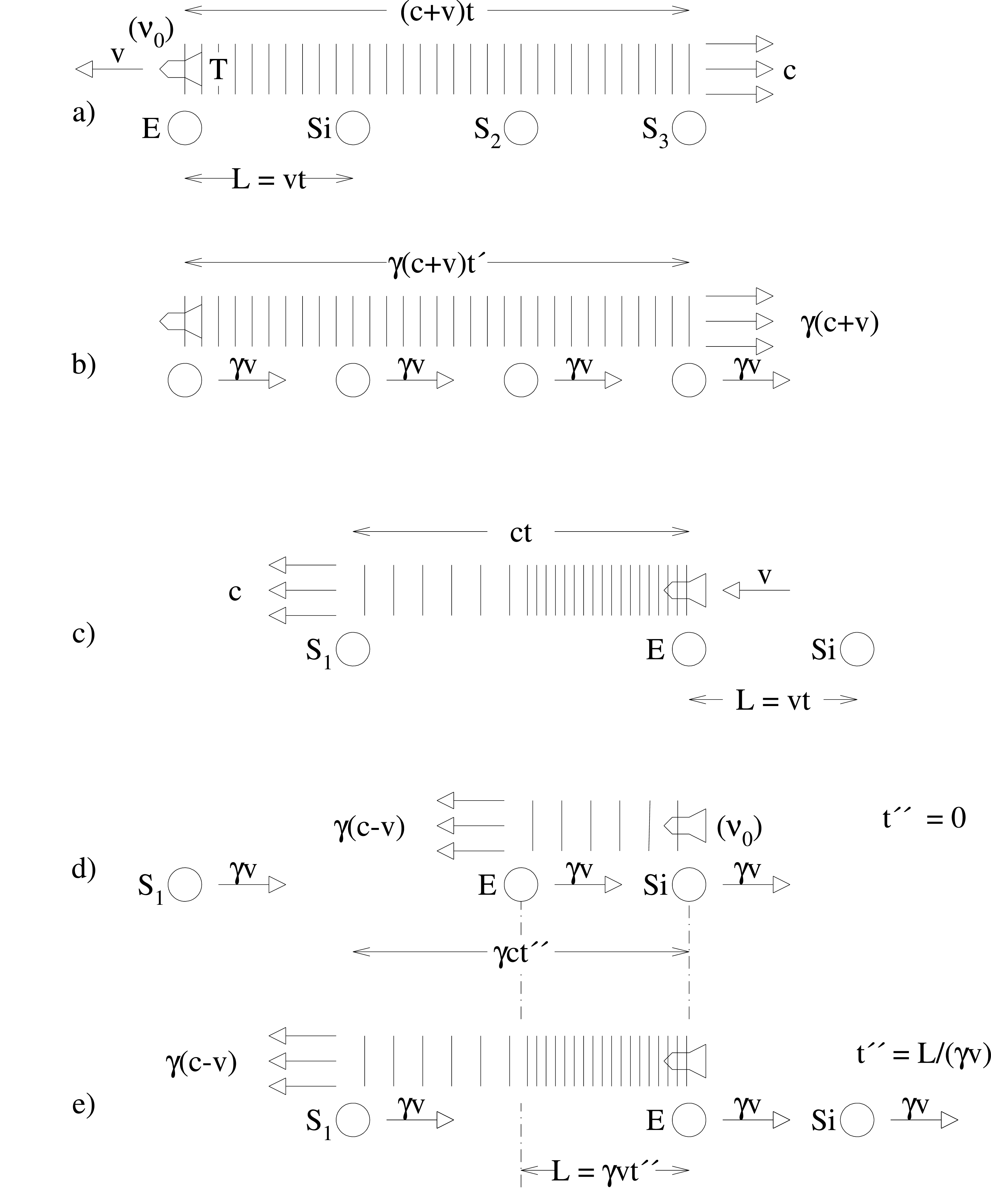}}
\caption{ {\em Spatial configurations of E,T,Si,S$_1$,S$_2$ and light signals during T's
          return journey. Clocks are reset to zero at the beginning of the journey. 
       a) In S, the rest frame of E,Si,S$_1$ and S$_2$, at the end
            of the return journey, showing light signals
         sent from the Earth at frequency $\nu_0$ (the same as Fig.~6a).
         b) in S'',the rest frame of T, at the end
            of the return journey, as calculated using
  postulates (iv) and (v). c) in S, at the end
            of the return journey, showing all the light signals, sent from T with time-dilated frequency $\nu_0/\gamma$ 
             that are received by E during the return journey (the same as Fig.~6c).
          d) in S'', at the beginning of the return journey, showing
             light signals sent by T at frequency $\nu_0$ during the outward journey, but not yet received by E, calculated
             as in b) above.
            e) in S'', at the end of the return journey, showing all light signals received by E during the
               return journey, calculated as in b) above. $v=c/2$, $\gamma = 1.155$. See text for discussion.}}
\label{fig-fig10}
\end{center}
\end{figure}
       \par The repetition frequency of the pulsed light signals in the thought experiment
        discussed in this section is defined as the reciprocal of the time interval between
        the passage of successive signals, whereas Langevin considered visible light signals of
        unspecified nature or continous electromagnetic (EM) waves transmitted and received
        by dipole antennae with length proportional to the wavelengths of the transmitted or
       received signals. When Langevin considers changes in the frequency or wavelength
       of the EM waves or optical light what are being discussed, at the fundamental level,
       are changes in the energy or momentum of the associated photons. In the above discussion,
        only the repetition frequency of the light pulses is considered from the viewpoint
       of space-time geometry. The energy, or frequency, of the photons constituting the
       light signals is nowhere mentioned. The transformations of frequency and velocity
       discussed above in terms of space-time geometry are now related to the changes of
       energy and momentum of the massive astronomical objects and massless photons
       under Lorentz transformations.
        \par The momentum-energy transformation equations for a physical
       object of Newtonian mass $m$, energy $E$ and momentum $p$ between the frames S and S
         are\footnote{Only motion parallel to the
          $\rx,\rx'$ axes is considered}:
         \ba
      p' & = & \gamma(p-\frac{\beta E}{c}) \\
      E' & = & \gamma(E-vp)
         \ea
   where
      \ba
        p & \equiv & \gamma_u m u  = \frac{Eu}{c^2} \\
        E & \equiv & \gamma_u m c^2
         \ea
     where $\gamma_u \equiv \sqrt{1-(u/c)^2}$ and
        \beq 
          E^2 = p^2c^2+m^2c^4
        \eeq
       Eq. (3.39) gives a relation between the velocity $u$, $p$ and $E$:
        \beq
        u = \frac{pc^2}{E}
        \eeq
      The PVAR of Eq. (3.9) is a consequence of (3.37), (3.38) and (3.42).
       Taking the ratio of (3.37) to (3.38):
      \beq
       \frac{p'}{E'} = \frac{p-\frac{\beta E}{c}}{E-vp} = \frac{1}{c^2}\frac{(\frac{p c^2}{E}-v)}
      {(1-\frac{v}{c^2}\frac{p c^2}{E})}
      \eeq
       so that, using (3.42)
        \beq
        u' = \frac{u-v}{1-\frac{uv}{c^2}}
       \eeq
       Using (3.44) to calculate the velocity of the Earth ($u =0$ in S) in the frame
        S' gives $u'= -v'=-v$ to be compared with the RRVTR prediction of Eq. (3.33), $v'= \gamma v$,
        which gives the observed velocity of E in S' in the thought experiment considered above.
        In fact the kinematically-derived formula (3.44) relates the configuration in
         the `base frame'~\cite{JHFSTP3,JHFRECP} of the `primary' experiment discussed above
         to that in the base frame (now S') of the `reciprocal' experiment in which E's clocks will
         appear to run slow to T, so that, after the round trip, it will be E that has aged
         less than T. This experiment is physically independent of that discussed above and shown in
         Figs. 5-10.        
         \par Since the photon is a massless particle, (2.42) and (2.43) give:
       \ba 
          E & = & pc \\
         u & = & c\left(\frac{pc}{E}\right) = c 
       \ea
       which is a first-principle kinematical derivation of Einstein's second postulate
       of special relativity, concerning the constancy of the speed of light~\cite{JHFHPA,JHFSTP1}.
        For photons therefore, (3.45) enables (3.38) to be written as:
        \beq
         E' = \gamma(1-\beta)E
        \eeq
        It then follows from the Planck-Einstein formula $E = h \nu$ (known at the time that Langevin wrote
        his paper) that
        \beq
        \nu' = \gamma(1-\beta)\nu =\sqrt{\frac{1-\beta}{1+\beta}}\nu
        \eeq
         showing the same frequency shift as the space-time analysis result of Eq. (3.34).
         \par It follows from the De Broglie formula (not known to Langevin in 1911):
         \beq 
            p = \frac{h}{\lambda_{DB}}
         \eeq
          and (3.45) and (3.47) that the De Broglie wavelength $\lambda_{DB}$ transforms as
                 \beq
        \lambda_{DB}' =\sqrt{\frac{1+\beta}{1-\beta}} \lambda_{DB}
        \eeq
          Using (3.45) to combine the Planck-Einstein and De Broglie formulae gives
        \beq 
        c = \nu \lambda_{DB}
         \eeq
         while from (3.48), (3.50) and (3.51)
          \beq 
            \nu' \lambda_{DB}' = \nu \lambda_{DB} = c
          \eeq
           The wavelength $\lambda_{DB}$ related to $p$ and Planck's constant, $h$, in Eq. (3.49  )
            is clearly a `kinematical' quantity to be contrasted. with the space-time geometric
            wavelength $\lambda_{geom}$ which is analogous to the separations $\delta$, $\delta'$
            of light pulses considered above. If  $\lambda_{geom}$ is defined in terms of the observed
            velocities and frequencies of photons in S and S' it follows from the geometry
            of Fig. 8 that
             \beq 
    \lambda_{geom}' \equiv \frac{u'}{\nu'} = \frac{1}{\nu}\sqrt{\frac{1+\beta}{1-\beta}}\gamma(1-\beta)c
               = \frac{c}{\nu} =  \lambda_{geom} 
            \eeq
            so that $\lambda_{geom}$  is a invariant quantity, like the
            intervals $\delta = \delta'$ in Figs. 9 and 10.
           \par It is clear that it is  $\lambda_{geom}$ and not  $\lambda_{DB}$ that is to 
             be associated with the phenomenological (from the standpoint of quantum mechanics)
             `wavelength of light' that occurs in the description of interference and diffraction
             phenomena in classical physical optics (the classical wave theory of light). This wavelength
                determines the phase of the wave, at any position, in terms of the spatial separation from 
               the source.
               Since both the phase and the spatial separation are Lorentz invariant the 
                wavelength must also be invariant. If only a single inertial frame is considered,
                in which photons are both created and destroyed, as is usually the case in physical
                optics problems~\cite{JHFAP} (3.51) and (3.53) give
                     \beq 
                     \nu \lambda_{geom} = \nu \lambda_{DB} =c
                    \eeq
                      so that $ \lambda_{DB} =\lambda_{geom}$ and both $\lambda_{geom}$ and $\lambda_{DB}$
                     are equivalent to the phenomological wavelength of the wave theory of light.
 
 \SECTION{\bf{Summary and closing remarks}}
   The main thesis of Langevin's essay ---that special relativity implied fundamental changes in
  the previous conceptual understanding of space and time--- was a valid one. However, the detailed discussion
   contains many self-contradictory arguments and physical misconceptions. Many of the former are specific to
    Langevin, whereas the latter, for example the RS and LC effects, are still the standard text-book interpretation
    of physical predictions of the space-time LT.
    \par Langevin asserts, on several occasions, that light and electromagnetic waves are vibrations of
     a mechanical aether, as envisaged by Faraday and later mathematically treated by Maxwell, in spite
     of the incompatiblity of such a model with Einstein's postulate of light speed constancy, which in other
    places is described by Langevin as the conceptual basis of special relativity. Langevin claims that,
     not only the Galilean space-time of Newtonian mechanics, but mechanics itself has been superseded by
    electromagnetic theory. This in spite of the knowledge of, and the first experimental verifications
     of, predictions of relativistic mechanics, derived from the LT, which already existed in 1911.

       \par Langevin states clearly that the length of a moving object is defined by simultaneous observation
        of the spatial positions of its ends, in the frame of reference in which the length measurement is
       made. The issue is then confused by considering the spatial sparation of non-simultaneous
      events (different times at which objects are dropped through a hole in the floor of a moving
        cart) that are in spatial coincidence for some observers and spatially separated for others.
        Langevin goes on to claim the TD effect of special relativity restores 
       a symmetry lacking in Galilean relativity where time intervals are observer-independent,
        since, in special relativity, both length and time intervals are observer-dependent.
        However, observer-dependence of length intervals occurs (in both Galilean and special relativity) only
        for non-simultaneous events. Length intervals defined by simultaneous measurement in some frame
        are invariant in both Galilean and special relativity, since, as demonstrated in Section 2,
       the LC effect is spurious.
        Thus, contrary to Langevin's assertion, the situation is more symmetrical in
        Galilean relativity where, (see Section 2) length intervals defined as $ t =  t' = 0$
      projections, and time intervals defined as $x' = 0$ projections, are both observer-independent,
       than in special relativity where only the latter intervals are observer-dependent.
      \par After this misleading discussion of the `relativity of length intervals', only applicable
          to non-simultaneous events, Langevin makes the important remark that events which are
          coincident both in space and in time, must be so for all observers, irrespective of their
          reference frame. This postulate, that may be called the `zeroth theorem of relativity',  
            leads in Ref.~\cite{JHFSTL} to the definition of a `corresponding epoch' in two 
            reference frames in relative motion.
   \par Langevin gives the usual, and incorrect, interpretation of the PVAR (3.9) in terms
      of the observation of the same light signal by observers in different inertial frames, instead of
        the transformation, between two inertial frames, of kinematical configurations corresponding to
        {\it physically independent} space-time experiments~\cite{JHFSTP3,JHFRECP}. Correct description
      of light signals in the travelling twin experiment, (see below) requires, instead, use
      of the RRVTR (3.10) to obtain consistent predictions for observations of both signal frequencies
      and signal numbers in the proper frames of the two twins.
       \par Langevin introduces the Lorentz-invariant categories of `space-like separated' and `time-like
  separated' pairs of events, which are defined as those that can (`time-like pairs') or cannot
        (`space-like pairs') be causally connected by light signals. The RS and LC effects
         described by Langevin as pertaining to space-like pairs are derived in Section 2 from the generic
        LT (2.4)-(2.6) (Eqs. (2.17), LS, and (2.18), RS, and Fig. 2). As previously pointed out in 
          Refs.~\cite{JHFLLT,JHFSTP1,JHFSTP2,JHFSTP3} the generic LT does not correctly describe
        a synchronised  clock at rest in S' at $x'= L$ when $L \ne 0$.  Using the correct
          LT for such a clock (Eq. (2.26) with $L'= L$, and Eq. (2.27)) it can be seen from Figs. 3 and 4
          that the RS and LC effects are spurious. Measured length intervals (defined as the spatial separation
          of a pair of simultaneous events) are therefore invariant, not only as stated by Langevin, in
          Galilean relativity, but also in special relativity. Time-like pairs, appropriate to observations
         of a single moving clock, are, in contrast, correctly described by the generic LT when the 
          clock is situated at the coordinate origin in S', leading to the prediction of the
         experimentally-confirmed TD effect (Eq. (2.10) and Fig. 1), as described by Langevin,
        for this case. 
        \par It is also demonstrated in Section 2 that the space and time coordinates that appear in 
          Eq. (2.17) (LC) and (2.18) (RS) cannot represent events on the world lines of synchronised
              clocks at fixed positions in S´, as is conventionally claimed to be the case. Following
        Ref.~\cite{JHFAJP1}, the projective geometry of the generic LT is discussed (see Table 1). It is 
         pointed out that only the $\rx' = 0$ and $\rx = 0$ projections (TD In a primary experiment and its
           reciprocal) correspond to physical effects, the predictions for the $\rx_0' = 0$ (LC) and 
         $\rx_0 = 0$ (LD) projections being invalidated by the presence of the additive constants 
         (see Eqs. (2.48)-(2.50)) needed (as first stated by Einstein) in order to correctly describe synchronised
         clocks at different spatial locations.
   \par Following Einstein, Langevin claims that
      the existence of a limiting speed of signal propagation, equal to that of light in free space, follows
      from causality and the existence of the RS effect. A signal travelling faster than the speed of light
      could establish a causal connection between two events with space-like separation. According
        to RS, the order of such events is observer-dependent and so some obervers could see
       the cause occur after the effect. Such an absurdity is avoided if no signal can travel
      faster than the speed of light. The flaw in this argument is that simultaneity and the
      time ordering of events are, in fact,  absolute, so that, as explained above, the RS effect
      is spurious. There is therefore no argument from causality forbidding
      signal propagation at superluminal speed.
       
         \par Langevin asserts that the impossiblity of instantaneous Newtonian action-at-a distance,
           claimed to be a consequence of causality and RS, gives support to Faraday's contact-interaction
           model of light propagation through a material aether, regardless of the incompatiblity
           (as pointed out by Einstein in Ref.~\cite{Ein1}) with the second postulate of special relativity.
        \par Before introducing the travelling-twin thought experiment, Langevin discusses the TD effect
          responsible for the differential aging which is predicted by special relativity to occur in
          the experiment. Langevin asserts, incorrectly, that the time interval between two events on the 
           world line of any `element of matter' is less for observers at rest relative to the element 
           than for any other observers whatsoever, concluding, also incorrectly, from this that acceleration
           is the cause of differential aging. Langevin's assertion is correct if, in the
            primary TD experiment discussed in Section 2, the `element of matter' is identified
           with the clock $\RC'_1$ which undergoes the TD effect. However if instead the `element of matter' 
          is identified with the clock $\RC_1$, while the observers are at rest in the frame S', in the same
           experiment, the latter will observe the time interval recorded by  $\RC_1$ to be greater than,
            not less than, the time recorded by  $\RC'_1$, relative to which they are at rest. Such an observation
            negates Langevin's assertion. In the (physically independent) reciprocal experiment
          where the moving clock $\RC_1$ is observed to run slow by observers at rest in S', Langevin's
           assertion is correct for $\RC_1$, but false for  $\RC'_1$. Specific counter
          examples~\cite{Halsbury,Marder,JHFSTP3} show that, contrary to Langevin's assertion, it is uniquely the 
          TD effect relating time interval observations in different inertial frames, not acceleration,
          that explains 
           differential aging in the travelling-twin experiment. 
          \par As a specific example of the TD effect, Langevin introduces `radioactive clocks' the activity
           (number of desintegrations per unit time) of which gives a direct measurement of the age of the clock.
              Langevin notes that such a clock, moving with a speed of 4000 km/s, will, after one mean lifetime,
             have aged less by one part in ten thousand than a similar clock at rest in the laboratory. 
              As pointed our elsewhere~\cite{JHFLLT,JHFSTP3}, consideration of two such identical, but
            spatially-separated radioactive clocks that undergo identical motion, makes evident the
            absurd and unphysical nature of the RS effect of standard special relativity theory 
             ---both clocks must show identical increments of differential aging relative to a clock
               at rest, i.e. the same TD effect, independently of their spatial positions.
             \par A detailed analysis of the travelling-twin experiment by the present author may be 
                      found in Ref.~\cite{JHFSTP3} where it is shown that the physical basis
                  of the differential aging effect is the RRVTR of Eq. (3.10), not LC, as in
                  the conventional interpretation. The analysis presented in the present paper 
                concentrates instead on space-time events and observed frequencies related to
               the exchange of light signals between the twins, not considered in Ref.~\cite{JHFSTP3}.
               In Section 3, calculations of geometrical configurations and frequencies are performed 
               on the basis of three different sets of postulates or assertions drawn from the following
               list:
                 \begin{itemize}
               \item[(i)] The generic LT (2.4)-(2.6)
               \item[(ii)] The PVAR Eq.~(3.9)
                \item[(iii)] Equality, at corresponding epochs in different inertial frames,
                             of recorded numbers of signals
                 \item[(iv)] The LT (2.39)-(2.40)
                  \item[(v)] The RRVTR (3.10)
                \end{itemize}
                The first set of calculations (A) the results of which are shown in Figs.~5 and 6, is based
                on the use of (i), (ii) for all massive objects and (iii). The second calculation (B)
                shown in Fig.~7 uses (i) and (ii) for all objects, including the photons constituting the
                light signals. The third set (C), shown in Figs.~9 and 10, uses postulates (iv) and (v). 
                The derivation of the RRVTR (v) from postulates (iii) and (iv) is shown in  Fig.~8.
                Postulate (v) is therefore
                a necessary consequence of (iii) and (iv), not an independent hypothesis.
                \par The postulate set A
                gives results for observed frequencies consistent with the relativistic Doppler shift 
                formula (Eq.~(3.49)) derived later in Section 3 from kinematical transformation formulae and the
                Planck-Einstein relation $E = h \nu$. In this approach, the physical explanation of
                the TD effect is LC in the frame S'. The observed numbers of signals are consistent with (iii).
                However, the velocities of the light signals transform between the frames S and S' according to
                classical formula: $u' =c \pm v$, in contradiction with the prediction: $u' =c$ of the PVAR.
                 \par If, as in the postulate set B, the PVAR is applied consistently to both massive objects
                and photons the calculated signal frequency in S' given by Eq.~(3.30) does not agree
                with the value given by the  Doppler shift formula, or that calculated in the frame S, assuming the 
                TD effect. Also, in this case the postulate (iii) is not respected by the results.
                At the end of T's journey, the first light signal sent from the Earth is at the position of 
                S$_2$ in S (Fig.~7c) and at the position of  S$_3$ in S' (Fig.~7d). Eleven signals from the Earth
                are recorded by T in S during the journey (Fig.~7c) to be compared with five signals in S' (Fig.~7d).
                \par For the calculation using postulate set C, length intervals are the same in the frames S and S',
                 the observed frequencies are as given by the Doppler shift formula, the postulate (iii) is respected
                and the physical basis of TD is the RRVTR. In this case there are no contradictions
                  either internally, as for postulate set A, concerning the PVAR, or with the fundamental postulate
                  (iii) and the
                  Doppler shift formula, as for postulate set B. Indeed, the spurious nature of the LC effect which
                  explains TD for the postulate sets A and B was previously demonstrated in Section 2 (see Figs.~3 and 4).
                  It is therefore suggested here that the correct physical description is that following from 
                  postulates (iv) and (v) (or, equivalently, from (iii) and (iv) ) as shown in Figs.~9 and 10.
               \par At the end of Section 3, the PVAR, Eq.~(3.9), is derived from the momentum-energy Lorentz
               transformation, and it is emphasised that such a kinematical transformation relates 
               kinematical configurations in the base frames of different space-time experiments, not
               observations of the same space-time events in the base and travelling frames of the same
               space-time experiment~\cite{JHFSTP3,JHFRECP}. 
            It is the latter misuse of the PVAR in the postulate sets A and B discussed
               above that leads to self-contradictory results or ones at variance with the fundamental
               relativistic postulates (iii) and (iv), or with the Doppler shift formula (3.49).
               The kinematically defined De Broglie wavelength of Eq.~(3.50) which transforms according
               to Eq.~(3.51) may be contrasted with the phenomenological geometrical wavelength of the classical
               wave theory of light, which is (see Eq.~(3.54)) a Lorentz invariant quantity. If, as is usually
               the case, all calculations are performed in a single reference frame where light is created,
               moves at speed $c$, and is destroyed, the De Broglie and geometrical wavelengths are the same.
               \par This review of Langevin's paper on special relativity and 
                  electromagnetism~\cite{Langevin}, almost a century after its publication, throws
                into sharp relief both the profound improvement in the understanding of fundamental physics
                  achieved during the 20th Century, and some important conceptual errors, 
               concerning physical space and time, which have persisted, uncorrected, throughout the 
                same period. With the present day knowledge of quantum mechanics, and in particular
               quantum electrodynamics (QED) Langevin's antithesis between mechanics and electromagnetism
               is essentially resolved, in favour of mechanics, by QED which is the (quantum) mechanics
                of interacting charged particles and photons. Indeed, many of the fundamental conceptions
               and equations of quantum mechanics can be derived by realising that a `plane electromagnetic
                 wave' is actually a parallel beam of monochromatic real (on-shell) photons~\cite{JHFEJP}.
                In this way, the Planck-Einstein relation $E = h \nu$ is derived by comparing the relativistic
                Doppler shift frequency, as derived by Einstein~\cite{Ein1} by postulating Lorentz invariance of the 
                 phase of an electromagnetic wave, with the kinematical formula (3.47)
                 describing the transformation of the energy of a massless particle (the photon).
                  A Lagrangian formulation of relativistic classical electrodynamics~\cite{JHFPS2} is obtained
                as the classical limit of QED. 
                Electric and magnetic fields and Maxwell's equations are derived by mathematical substitution
                in the equations of this theory which is formulated without introducing any {\it a priori} `field' 
                concept. QED also predicts that force fields, with $1/r^2$ spatial dependence,
                 mediated by the exchange of space-like virtual photons, are transmitted instantaneously~\cite{JHFPS2},
                consistent with the results of a recent experiments~\cite{JAP1,JHFNRQED,Frascexpt}, as well as original
                observations
                of Hertz at small transmitter-receiver separations~\cite{RSRFF,Buchwald,Hertz}. The classical wave
                 theory of light follows from the QED description of space-time events connected by the paths
                 of real photons~\cite{FeynQED,JHFAP}.
                 \par The essential idea which Langevin attempted to communicate in Ref.~\cite{Langevin}, in simple
                 language, without the use of mathematics, is both important and correct ---a revolutionary change
                in the commonsense concept of time, required by special relativity, and exemplified in a graphical
                 way by the differential aging that occurs in the travelling-twin thought experiment ---the travelling
                twin ages less than the stay-at-home one. However, there were many errors, some purely logical
                (self-contradictory statements) others of a conceptual nature, in Langevin's account, which it
                 is instructive to divide into two categories:
                 \begin{itemize} 
                 \item[(a)] Those that require a deeper and novel (for 1911) understanding of physics
                              for their resolution
                  \item[(b)] Those that require, for their correction, only the knowledge of physics
                              which already existed in 1911
                   \end{itemize} 
                   To completely resolve the apparent antithesis between classical mechanics and classical
                   electrodynamics emphasised by Langevin, two ingredients were necessary: firstly, 
                      relativistic mechanics which, following the work of Einstein~\cite{Ein1},
                    Planck~\cite{PlanckRM}, Poincar\'{e}~\cite{Poincare} and Minkowski~\cite{Mink} 
                 was aleady well-known at the time
                 Langevin's paper was written, and, secondly, QED, yet to be discovered at that time.
                 Essentially all of Langevin's other logical errors and misconceptions fall into the category
                 (b). The evident contradiction between the mechanical-aether model of light propagation
                  by contact interactions, asserted to be physically correct by Langevin, and Einstein's 
                second postulate of special relativity
                is
                   not remarked upon. Unnoticed also is Einstein's related remark on the superfluous nature
                  of the aether concept. Following Einstein~\cite{Ein1} and assuming the existence of the RS
                  effect  Langevin adduces an argument that derives a maximum signal speed 
                  equal to that of light from the requirement of causality (that no observer can see an effect
                    occur before its cause). This conclusion is invalidated since, as explained 
                    in Section 2 above, RS does not exist when LT equations describing correctly
                    synchronised clocks different spatial locations are used.
                  \par Langevin does not discuss the physical mechanism underlying TD and the differential
                   aging effect, but since he asserts the existence of the LC effect and that the observed 
                   velocity of a light signal in a space-time experiment transforms according to the PVAR,
                   the spatial configuration
                   in the rest frame, S', of T, of the light signals sent from the Earth, is that shown in
                   Fig.~7d. If Langevin has actually drawn and inspected this figure in 1911 (as he could
                   well have done) he could have noted the following features:
                   \begin{itemize}
                     \item[(I)] The number of signals received by T from the Earth is different
                     in S and S'. 
                     \item[(II)] At the end of the journey, the first light signal sent from the
                                 the Earth is in spatial coincidence with S$_2$ in S and with S$_3$ in S'.
                     \item[(III)] The frequency of the signals recorded by T in S' (given by Eq.~(3.30))
                                  is different to the value, as correctly calculated by Langevin
                                   for the configuration in Fig.~7c in S, for $\gamma = 100$,
                                   on assuming the TD effect, (Eq ~(3.11)).
                     \end{itemize}
                      (II) is in violation of the postulate (iii) stating the frame-independence of the
                     number of signals recorded by any observer.
                        \par If Langevin had noticed the contradictions apparent in Fig.~7 he might have
                       been lead to reflect on the physical characteristics of clocks, which,
                       no different to the present day, had, in 1911, a setting (epoch) completely under the
                         control of the experimeter using  them, and which ran at a rate determined by the
                       physics of the clock mechanism, not controlled by the experimenter~\cite{JHFCRCS}. If Langevin
                         had read carefully Einstein's 1905 paper~\cite{Ein1} and noted the important
                      passage concerning the additive constants in the LT to correctly describe 
                      synchronised clocks at different spatial locations, he might have obtained
                      Eqs.~(2.48)-(2.50) and realised that the RS and LC effects (but not TD) are
                       spurious, due to a simple confusion of the offset of a desynchronised clock 
                       with a physical time interval~\cite{JHFCRCS}. 
                       \par What the present author finds amazing is that, if, in the discussion of Fig. 7 just given, 
                       the  word `Langevin' is replaced by `any physicist' it remains valid and relevant today
                         and was so throughout the 20th Century!
                      As shown in Table 1 above, RS and LC are indeed the
                     rigorous mathematical predictions derived from certain projection operations on the 
                      generic LT (2.4)-(2.6) (see Table 1). The problem is, that these equations
                       do not, as has been universally assumed to be the case, describe a synchronised
                       clock at an arbitary spatial position. It is simply  a question of
                        using equations that correctly describe the actual (but arbitary
                          and completely controllable) initial conditions of the problem.
                          \par In conclusion, Langevin's conceptual errors of the category
                           (b) are the same ones that are to be found in all text books and all the pedagogical
                           literature on the theory of special relativity (with, to his present best knowledge,
                          the exception only of the work of the present author since 2004)
                           since its inception in 1905.
\pagebreak

\end{document}